\documentclass[journal]{vgtc}                     

\onlineid{1040}

\vgtccategory{Research}
\vgtcpapertype{application/design study}

\title{Co-Designing Dynamic Mixed Reality Drill Positioning Widgets: A Collaborative Approach with Dentists in a Realistic Setup}

\author{
\authororcid{Mine Dastan}{0000-0003-0555-155X},
\authororcid {Michele Fiorentino}{0000-0003-2197-6574},
\authororcid{Elias D. Walter}{0000-0003-4802-2279},
\authororcid{Christian Diegritz}{0000-0002-1034-6844},\\ 
\authororcid {Antonio E. Uva}{0000-0001-7271-6137},
\authororcid{Ulrich Eck}{0000-0002-5322-4724}, and
\authororcid {Nassir Navab}{0000-0002-6032-5611}
}

\authorfooter{
 \item
	Mine Dastan, Michele Fiorentino and Antonio E. Uva  are with the Polytechnic University of Bari.
	E-mail: mine.dastan@poliba.it
  \item
   	Elias Walter and Christian Diegritz are with the Ludwig Maximilian University of Munich.
  
  \item Ulrich Eck and Nassir Navab are with the Technical University of Munich.
}

\abstract{%
  Mixed Reality (MR) is proven in the literature to support precise spatial dental drill positioning by superimposing 3D widgets. Despite this, the related knowledge about widget's visual design and interactive user feedback is still limited. Therefore, this study is contributed to by co-designed MR drill tool positioning widgets with two expert dentists and three MR experts. The results of co-design are two static widgets (SWs):  a simple entry point, a target axis, and two dynamic widgets (DWs), variants of dynamic error visualization with and without a target axis (DWTA and DWEP). We evaluated the co-designed widgets in a virtual reality simulation supported by a realistic setup with a tracked phantom patient, a virtual magnifying loupe, and a dentist's foot pedal. The user study involved 35 dentists with various backgrounds and years of experience. The findings demonstrated significant results; DWs outperform SWs in positional and rotational precision, especially with younger generations and subjects with gaming experiences. The user preference remains for DWs (19) instead of SWs (16). However, findings indicated that the precision positively correlates with the time trade-off. The post-experience questionnaire (NASA-TLX) showed that DWs increase mental and physical demand, effort, and frustration more than SWs. Comparisons between DWEP and DWTA show that the DW's complexity level influences time, physical and mental demands. The DWs are extensible to diverse medical and industrial scenarios that demand precision.
  
}

\teaser{
  \centering
  \includegraphics[width=\linewidth]{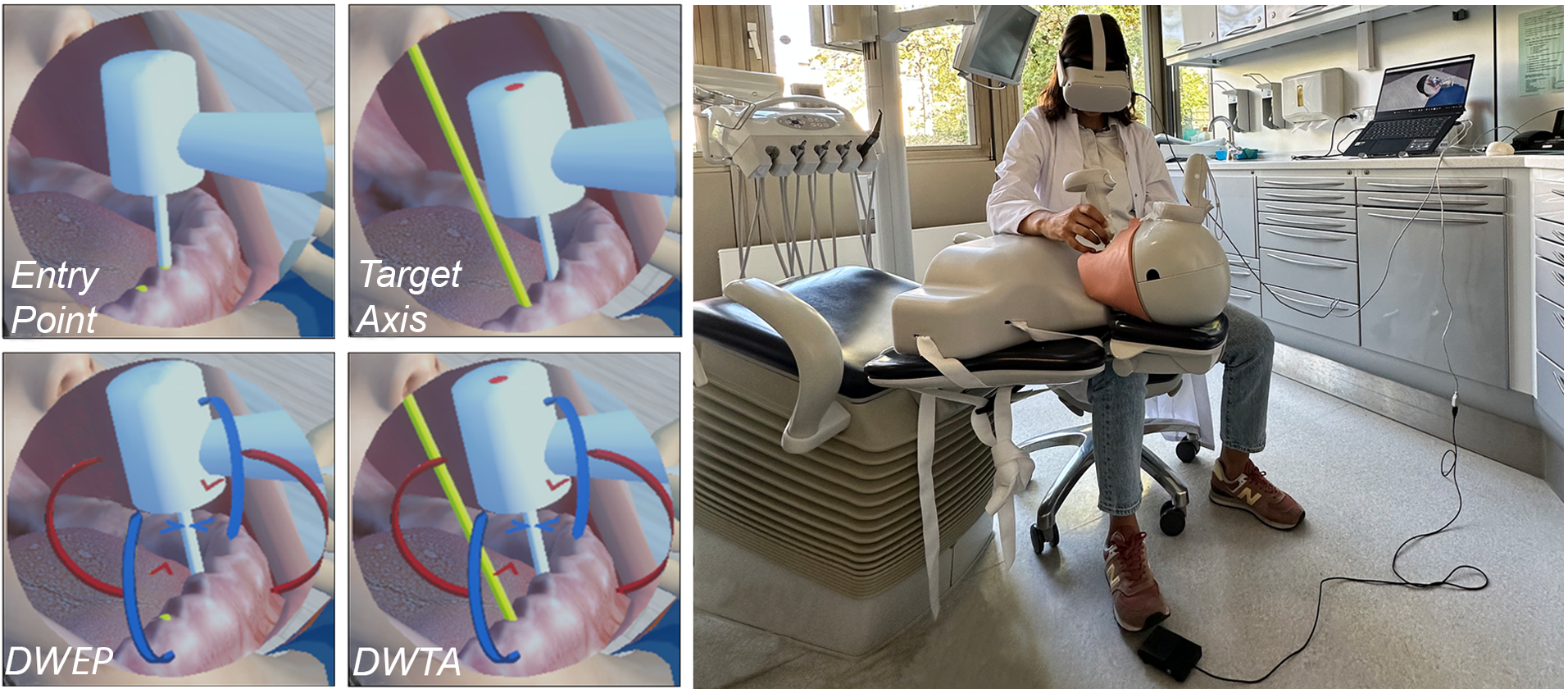}
  \caption{
  Co-designed mixed reality drill positioning widgets: Entry Point, Target Axis, Dynamic widget and entry point (DWEP), Dynamic widget and target axis (DWTA), compared in a realistic setup in the dental room: virtual magnifying loupe (left), tracked phantom and virtual patient, foot pedal, laptop and headset (right).}
  \label{fig: teaser}
}

\graphicspath{{figs/}{figures/}{pictures/}{images/}{./}} 

\usepackage{tabu}                      
\usepackage{booktabs}                  
\usepackage{lipsum}                    
\usepackage{mwe}                       

\usepackage{mathptmx}                  
\usepackage{marvosym} 
\usepackage{amssymb} 
\usepackage{booktabs} 
\usepackage{amsmath} 
\usepackage{amsfonts,mathtools} 
\usepackage{enumitem} 
\usepackage{fancyhdr}
\begin{document}

\firstsection{Introduction}

\maketitle

Precise drill positioning in dental implantology is one of the most challenging skills to master, requiring both time and spatial ability \cite{wang2023exploring,spille2022learning,cassetta2017much,cassetta2020there}. Pre-operative scans and software are utilized to calculate accurately the drill's 3D position and path axis \cite{block2017implant,takacs2023advancing,schubert2019digital}. The drill must be positioned in space by precisely coordinating 5 degrees of freedom (DoF), three positional components, and two rotational components (rotation around the y-axis, the local reference frame of the drill along the drill bit, is not crucial). This task involves operating in an environment with a moving patient and reduced visibility (shadows, fluids, and mists) that is physically and mentally demanding for the dentist \cite{ohlendorf2017constrained,park2015analysis,raikar2017factors}. Factors such as a narrow mandibular ridge can further increase implantation difficulty. \cite{galindo2017influence,bertollo2011drilling,yari2024risk}. Correct drill positioning and angulation are imperative not only for successful aesthetic and functional prosthodontic restorations but also to prevent irreversible damage to adjacent structures caused by inaccurate drilling. These include the alveolar nerve, teeth, or the sinuses, all posing risks of long-term complications and implant failure.\cite{mistry20213d,greenstein2015nerve,jung2008systematic}.

A proven solution is applying custom-made 3D physical templates \cite{schneider2009systematic}, which are expensive, disposable, and time-consuming to manufacture. On the contrary, current software-based drill positioning assistance \cite{dure2021first,chackartchi2022reducing,wu2024impacts,markovic2024considerations} relies on 2D screens, introducing the issues of divided attention, cognitive stress, and manual skill errors. Mixed Reality (MR) demonstrated to support drill positioning by superimposing mouth-referenced computer-generated 3D assets like a target axis path \cite{fahim2022augmented,pellegrino2021dynamic,farronato2019current}.

Spatial relationships in virtual environments present challenges to accurately perceive the depth and spatial orientation of manipulated objects \cite{zhai1993human,el2019survey,yeo2024entering,besanccon2021state,mendes2019survey,kaplan2021effects}. 3D virtual assistive elements, also known as widgets, generally convey information about precision tasks, including positioning. These widgets, defined as an "encapsulation of geometry and behavior used to control or display information" \cite{conner1992three} assist users in the interaction with scene elements \cite{henderson2011augmented,volmer2023multi,mendes2016benefits,mine1997moving}.

In medical scenarios, the user interface (UI) design must prioritize intuitiveness and comprehensibility to enhance usability, patient safety, and decrease task load \cite{muhler2009medical,shaalan2020visualization,ma2023visualization,preim2007visualization}. Potential UI elements are MR drill positioning widgets (MRDPW), which positively impact guidance and procedural tasks \cite{ma2019augmented,lin2015novel,dastan2022gestalt}. Most of the MRDPW in literature are pseudo-static \cite{kivovics2022accuracy,farronato2019current,tao2024comparison}, and dynamic designs have been explored recently. Nevertheless, the literature currently lacks a comparison of different MRDPWs.
The visual elements in UIs are often not optimized for a good user experience and end users are rarely included in the iterative development phase. Additionally, the widget validation of different widgets in the literature is incomparable and mostly not validated by realistic setups.
Considering tracking is not the major problem in the next-gen MR interfaces, we focus here on end-user experience, visualization, and efficacy. We use a co-design method that involves the end user in the process \cite{el2020can,bird2021generative,busciantella2024research} and that has been proven to be effective in meeting users' needs \cite{slattery2020research,harrison2022implementing}. Thus, the main goal of this paper is to co-design MRDPWs for achieving maximum performance with several research questions:
 \begin{itemize} [noitemsep]
     \item \textit{R$Q_1$: "What is the co-design outcome regarding widget design, their evaluation criteria and conditions?"}
     \item \textit{R$Q_2$: "Which condition is the best-performing and preferred widget in a realistic setup?"}
     \item \textit{R$Q_3$: "Are dynamic widgets more precise than static widgets?"}
     \item \textit{R$Q_4$: "Does precision impact other variables such as time and task load?"}
     \item \textit{R$Q_5$: "Do age and gaming experience influence precision?"}
 \end{itemize}

We evaluated co-designed conditions regarding precision, time, and task load using qualitative and quantitative measurements in a realistic setup. Lastly, we analyzed the dentist's opinions and preferences and present an overview of MRDPWs with key takeaways.

\section{Related Works}
Firstly, we analyzed the UIs and utilized widgets and setups for tool positioning in recent articles to generate an overview of existing methods, targeting primarily papers on dental implantology. The topic's significance is driven by the constantly increasing number of implants performed \cite{elani2018trends} and the need for assisted tools essential to preserving and enhancing patient care quality.
A commonly applied concept involves superimposing pre-operatively planned data over the surgical area with or without an auxiliary axis \cite{zhu2017novel,pietruski2019supporting}. This virtual widget is typically represented as a thin yellow trajectory line, which can be visually challenging. The visual feedback is only reinforced through a color change of the widget, which may not be sufficient for 5 DoF precision or accessible for each user.

A recent study by Ferronato et al. \cite{farronato2023novel} developed a novel tablet-based marker-less AR system using iPad Pro 2020 (Apple, CA, USA) for endodontic treatments; the UI consists of static reference points, so-called points of interest (POI), and red-colored target points for cavity position. A red central line is defined as a target axis for orientation. The system also inserted mandibular and maxillary CBCT volumes into the system.  Two dentists experimented on phantom and 3D-printed models with 90 drilling tasks. The results indicated that the positional error was 0.51\,mm and 0.77\,mm with 8.5\,° mean angular deviation.

 The work of Dastan et al. \cite{dastan2022gestalt} implemented the Augmented Collimator Widget (ACW) inspired by Gestalt theory. The work features a 3D widget comprising 2D elements for positional and rotational alignment. The ACW utilizes the visual error amplification method, in which the components separate when the tool is distant from the target and collimate as the tool approaches the target. The components disappear once the tool surpasses the threshold. The authors compared ACW with the golden standards (GSW) with 30 participants using Oculus Quest 2 (Meta, CA, USA). Yet, participants were not dentists and were not provided with a phantom. The results demonstrated that ACW performed better than the defined golden standards in AR simulation in VR. ACW outperformed in positional accuracy (2.24\,mm ± 1.42 for ACW vs. 2.72\,mm ± 2.75 for GSW) and rotational accuracy (5.03\,° ± 3.13 for ACW vs.  9.54\,° ± 5.77 for GSW). On the other hand, ACW resulted in an increased task time of 5.19\,sec ± 4.91 (ACW) vs. 2.14\,sec ± 1.81 (GSW).

 Another approach from Ma et al. \cite{ma2019augmented} involved implementing an AR tool navigation system using a target axis,  tool axis, and implant path superimposed on the surgical area through an IV overlay device. The authors conducted experiments on a phantom and a volunteer, with only one experienced dentist involved. The non-guided method (based solely on the dentist's experience) was compared with the AR system, and results demonstrated that the AR method performed better, showing positional errors 1.25\,mm vs. 1.63\,mm; rotational error = 4.03\,° vs. 6.10\,°. However, it was noted that using bulky trackers created discomfort for the volunteers in the AR setup.

Alternatively, Song et al. \cite{song2018endodontic} implemented the first head mounted display(HMD)-based AR prototype for endodontic procedures. Their work utilizes visual and audio cues during the interaction. The visual cues include a thin cylinder representing the virtual drill tool axis and two flat dynamic disks (one on the tooltip and one on the top of the tool). The disks indicate depth, distance, and orientation between the tool and target through color and size variation. When the tool is distant, the disks appear red and large.
Additionally, text feedback provides numerical errors (for position, rotation, and depth) in the dentist's field of view. The authors experimented on a scaled tooth model using HoloLens 2 (Hololens, Microsoft, USA) to evaluate their system with a series of tool target positioning. The positional error ranged from 3.6\,mm to 32.2\,mm, and the rotational error ranged from 2.15\,° to 45.10\,°. The authors claimed that their system reduces the decision time, and integrating visual cues makes the system usage more intuitive.

The work of Lin et al. \cite{lin2015novel} proposes an AR interface that requires the traditional surgical template. The UI includes the drill axis, tooltip, implant path, and mandibular nerves. The study conducted a phantom experiment with only one experienced periodontist using Sony HMZ-T1 (Sony Electronics Inc., CA, USA) HMD goggles. The research evaluated the accuracy of their systems with the pre-planned data. The findings demonstrated that the implant placement deviated less from the planned position with their AR system, showing a positional error of 0.50 ± 0.33\,mm and a rotational error of 2.70 ± 1.55\,° for mandibular implant placement.

For instance, the research by Wang et al. \cite{wang2014augmented} elaborated a phantom experiment for marker-free image registration for dental surgery. The virtual Drill Tool tip and the Trajectory Axis are superimposed on the phantom model using a projector-based AR system using a 3D image overlay device. The work calculates the registration error of 0.71 ± 0.27\,mm.

Moreover, Katic et al. \cite{katic2010knowledge} developed a context-aware AR dental implant surgery system. The authors implemented two visualization types: A direct overlay over the surgical field and a static visualization of information in a fixed position. A dentist superimposes two thin cylinders, the target axis and the drill axis, using the color-coding mechanism (red-green). The authors evaluated the registration accuracy of their AR system using HMD in a cadaver experiment. The findings demonstrated that the positional error significantly improved with their calibration system to less than 2.5\,mm.
\begin{figure*} [t]
    \centering
    \includegraphics [width=\textwidth]{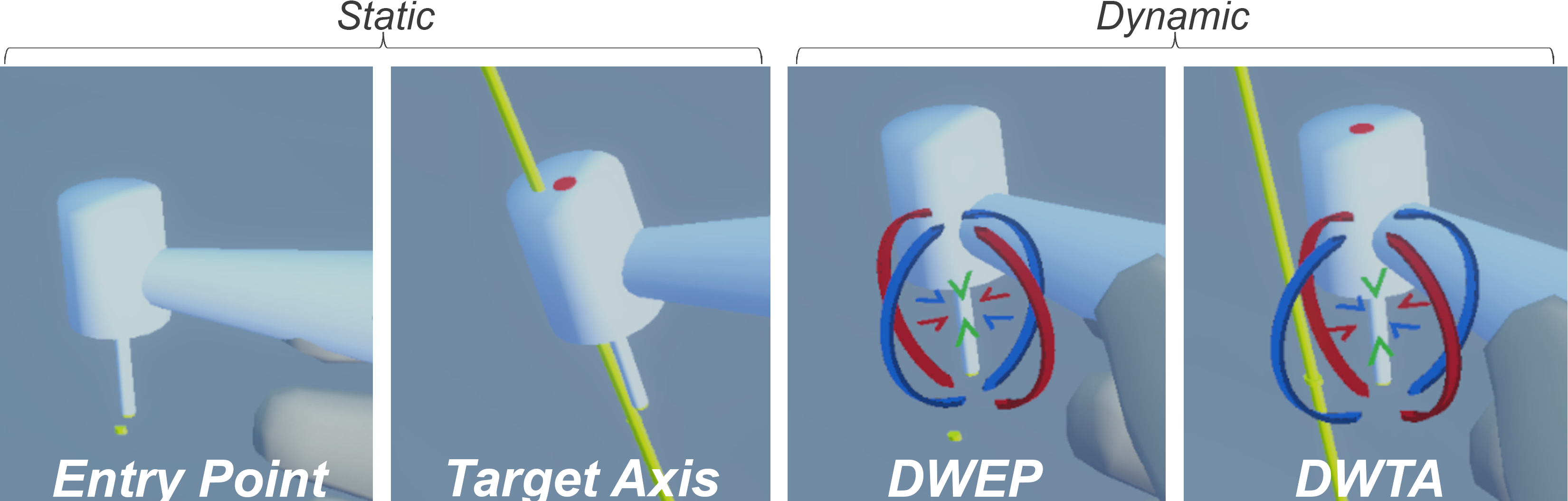}
    \caption{Static and Dynamic MRDPWs demonstration of only assistive virtual elements; entry point, target axis, DWEP dynamic widget with entry point, DWTA dynamic widget and target axis.}
    \label{fig:onlywidgets}
\end{figure*}
More recently, Tao et al. \cite{tao2023comparison} conducted a comparison between AR-based (ARN) and dynamic screen navigation (DSN) \textit{in vitro} studies using phantoms. Their system utilizes the entry point and trajectory axis to indicate the target. A total of 242 implants were placed by only one surgeon wearing HoloLens 2. The results were 1.31 ±0.67\,mm (ARN) vs.1.18 ± 0.59\,mm (DSN) for entry point deviation and 3.72 ±2.13\,° (ARN) vs.3.1 ±1.56\,° (DSN) rotational error. The findings indicated that both conditions performed similar positional errors; however, the study claimed that their AR-based navigation system yielded a higher angular deviation.

The existing body of literature provides valuable insights, while our research objectives aim to contribute to this goal by further refining and advancing the state-of-the-art. Most AR dental tool navigation systems primarily focus on calibration, registration, and real-time tracking issues \cite{tao2023comparison,ma2019augmented, lin2015novel, wang2014augmented}.  Our research objectives are aligned to pursue enhanced precision guidance by analysis of different visualization methods.

In conclusion, this paper contributes by;
\begin{itemize} [noitemsep]
    \item Co-design of MRDPWs with three MR experts and two dentists.
    \item As a result of co-design, the following widgets are implemented. Two static: \textit{\textbf{Entry Point}}, and \textit{\textbf{Target Axis}} inspired by literature. Two novel non-linear behavior Dynamic Widgets: \textit{\textbf{Dynamic widget with Entry Point}} and \textit{\textbf{Dynamic widget with Target Axis}}.
    \item Evaluation of MRDPWs with dentist subjects of various experience levels in a co-designed realistic setup.
\end{itemize}

\section {Widgets Co-Design with Dentists}
We followed the Co-design methodology proposed by  \cite{freudenthal2011collaborative} which specifically supports multidisciplinary medical collaboration by a user-centered iterative development process. We construct a co-design group consisting of five experts: two dentists with 15 and four years of experience in endodontology and three  MR experts with 26, 15, and five years of experience. Over four months, we performed one-to-one interviews, live demos, and focus groups \cite{o2018use}.

Physical and VR scenarios for the dental room simulations were implemented during the co-design. This approach allowed us to quickly assess altered widget design cycles, as a full AR setup is sensitive to trackers' error and latency.
Firstly, the experts discussed the MRDPW problem to evaluate an unbiased clinical dentist's perspective regarding visualization. Afterward, reproduced widgets from the literature were demonstrated. Furthermore, feedback for improvement and user experiment details were iteratively gathered and implemented with adjusted functionalities, particularly widgets' type, behavior, form, size, and material.

\subsection{Co-Design Results}
 At the end of the co-design phase, we answer our R$Q_1$. As an outcome, we implemented the following MRDPW conditions in a realistic setup (\cref{fig:onlywidgets}) with distinctive attributes (static vs dynamic) meeting the diverse needs of dentists when operating, such as visibility, usability, and design preferences.
  \begin{figure} [ht]
  \centering
  \begin{subfigure}[b]{0.45\columnwidth}
  	\centering
  	\includegraphics[width=\textwidth]{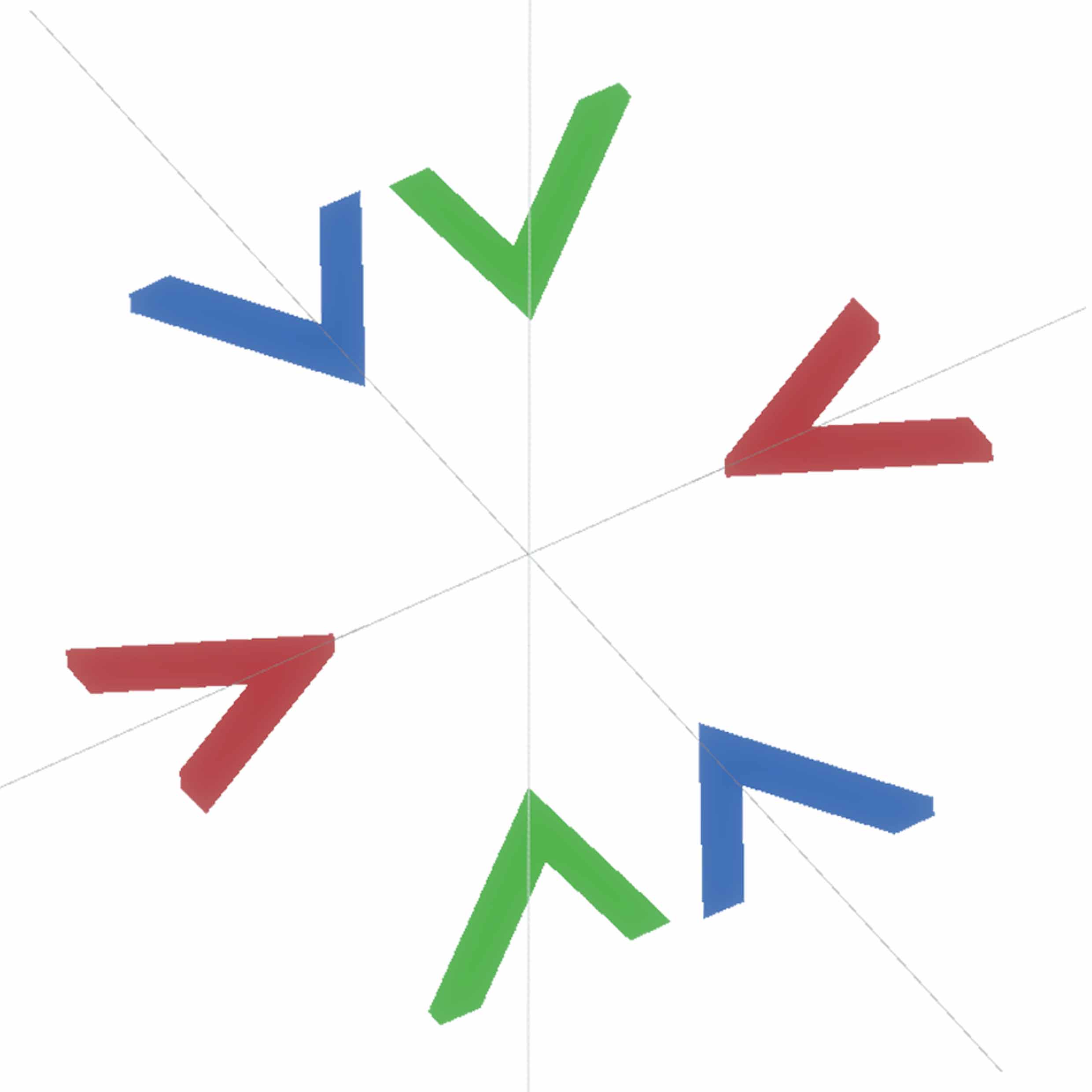}
   \caption{The positional components x, y, and z axis represented as 3D "\textbf{V}".}
  	\label{fig: DW V}
  \end{subfigure}%
  \hfill%
  \begin{subfigure}[b]{0.45\columnwidth}
  	\centering
  	\includegraphics[width=\textwidth]{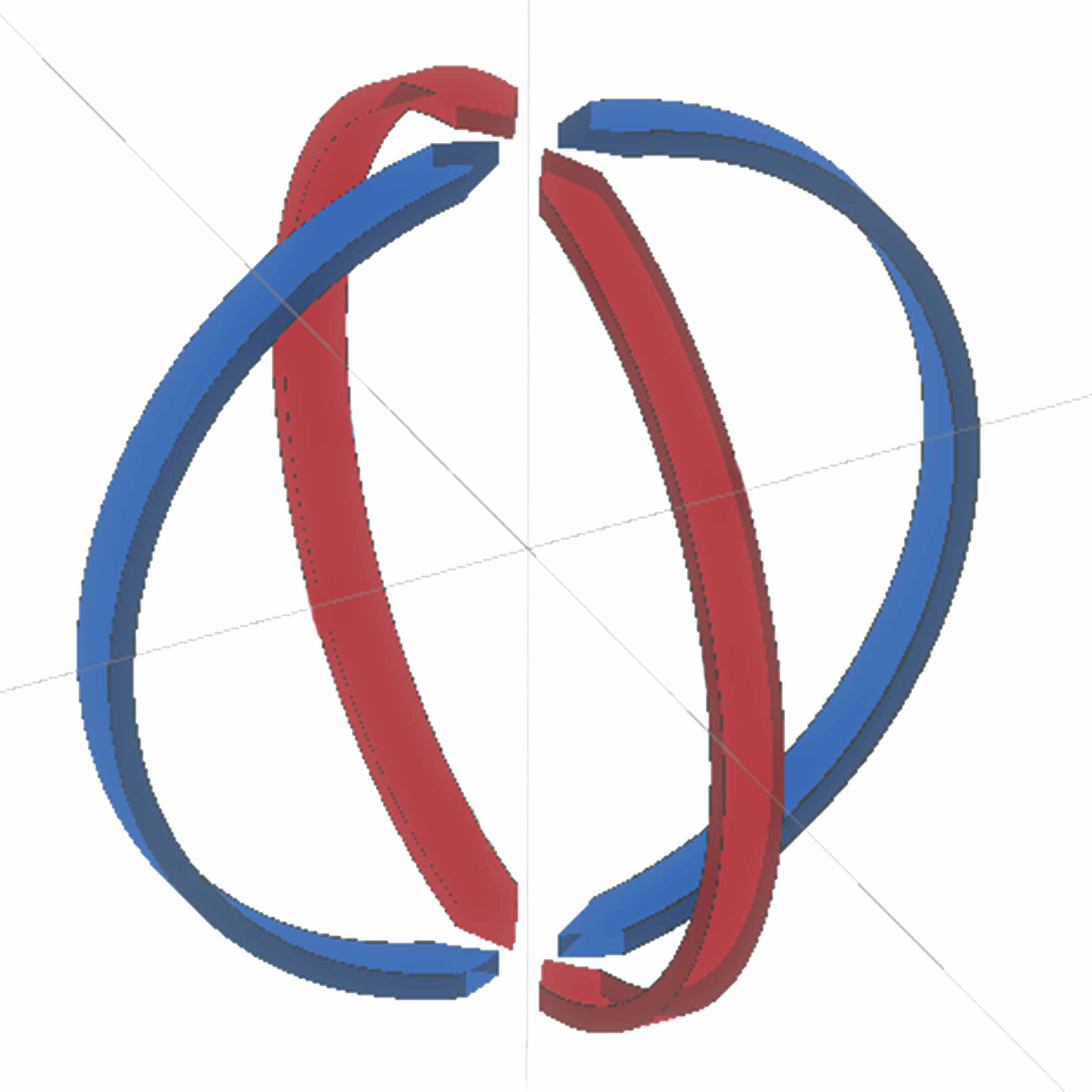}
   \caption{The rotational components, x and y axis represented as 3D "\textbf{(}".}
  	\label{fig: DW rot}
  \end{subfigure}%
  \\%
  \begin{subfigure}[b]{0.45\columnwidth}
  	\centering
  	\includegraphics[width=\textwidth]{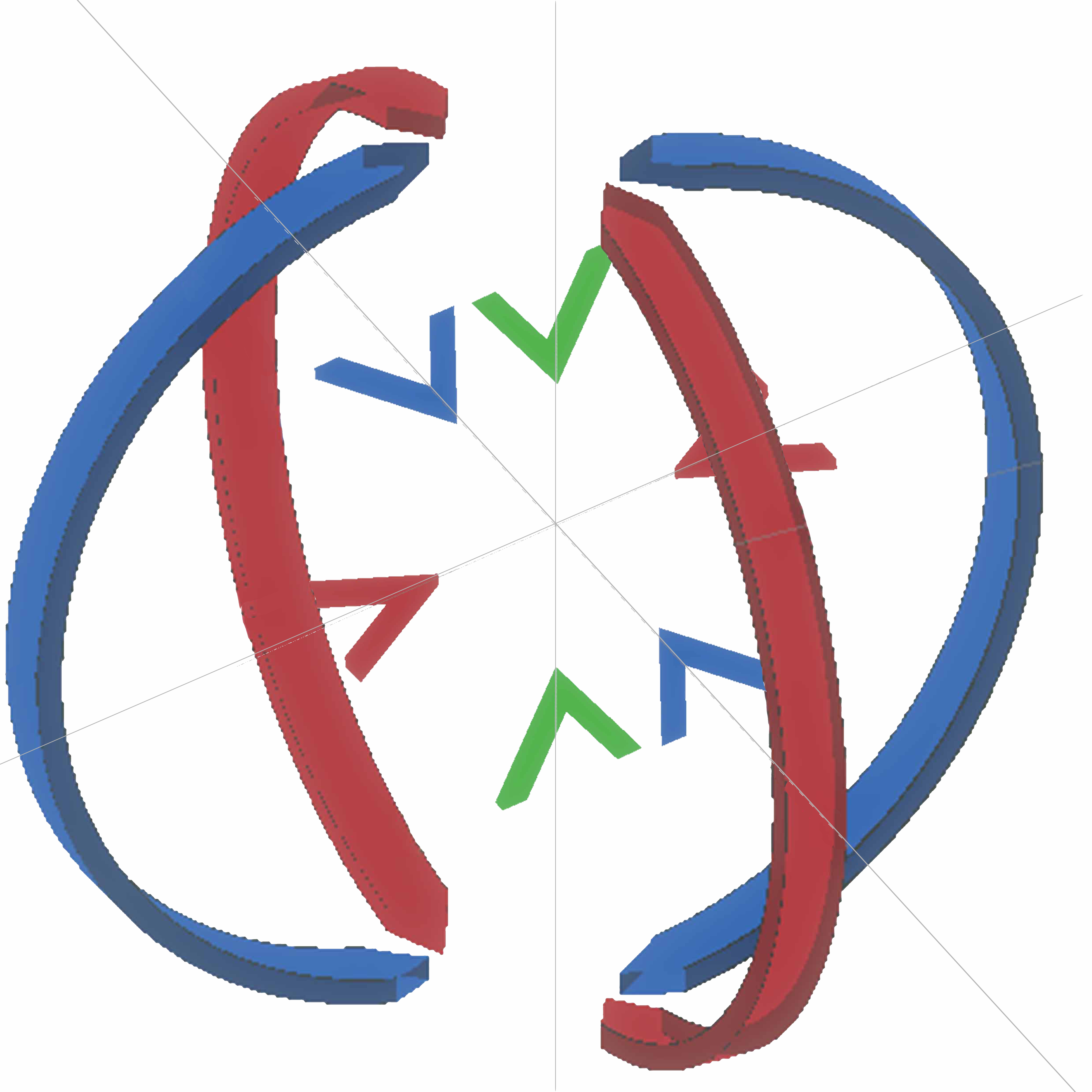}
   \caption{Tool is away from the target, forms are not collimated.}
  	\label{fig: dw non collimated}
  \end{subfigure}%
  \hfill%
  \begin{subfigure}[b]{0.45\columnwidth}
  	\centering
  	\includegraphics[width=\textwidth]{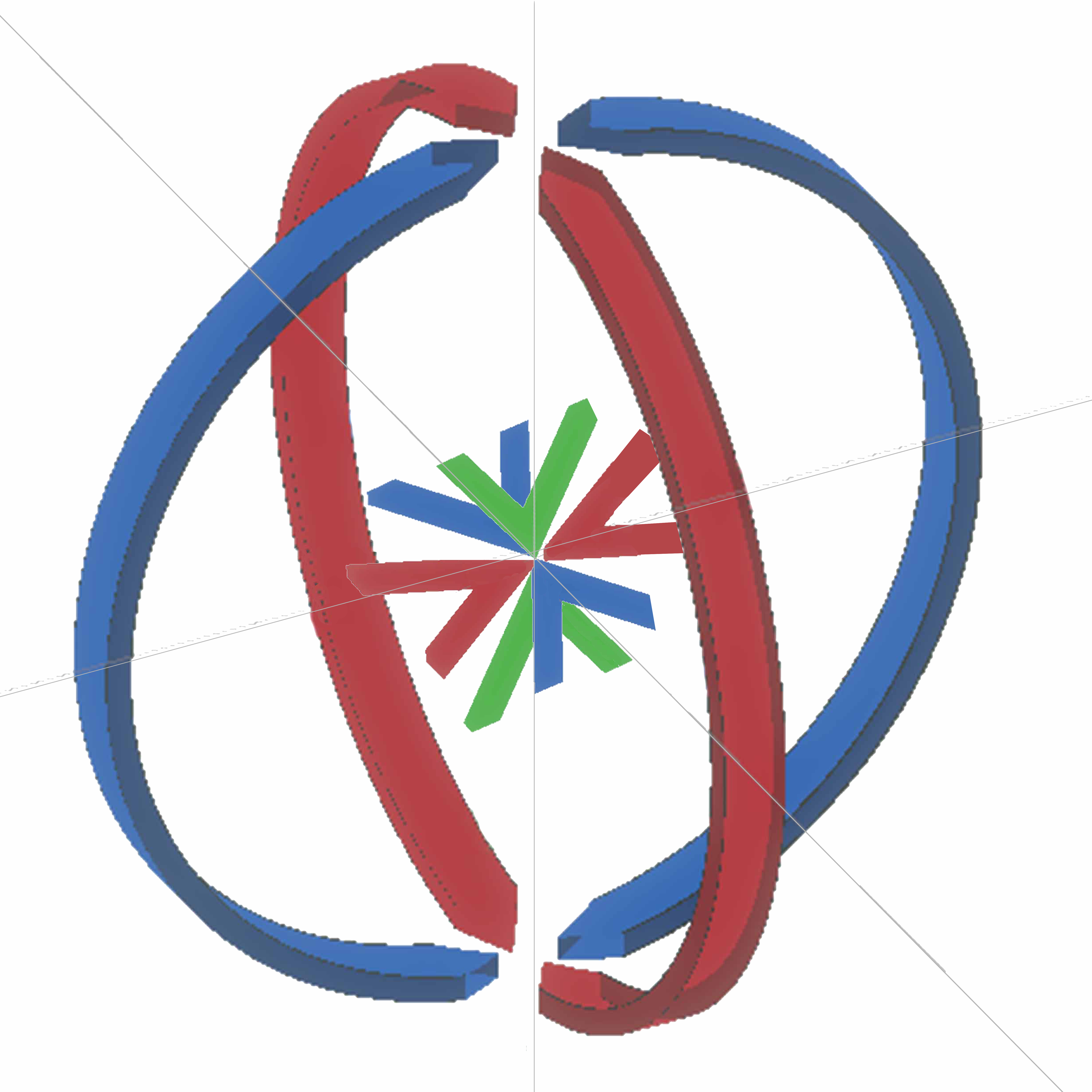}
   \caption{Tool is on the target, forms are collimated.}
  	\label{fig: dw COLLIMATED}
  \end{subfigure}%
  
  \caption{Compositions of DW: positional and rotational components (a, b), the tool is far from the target, forms are in the periphery and away from each other (c),the tool is on target, forms are nearby (d).}
  \label{fig: Total DW}
\end{figure}
Two MRDPWs are designed as static widgets and two as dynamic conditions \cref{fig:onlywidgets}.

We use Unity Platform \cite{Unity} with a world-fixed dextrorotary coordination system with z (blue) pointing forward to the initial user position, x (red) pointing to the horizontal, and y (green) pointing up on the vertical axis.
An unexpected outcome was the discovery that the industry standards of dentistry often support the right hand for the drill tool independently from the user's dominance. Therefore, we designed our widget to be right-handed, considering that most left-handed dentists also use their right hand for dental procedures.
 
\subsubsection{Static Widgets (SWs)}
\paragraph{\footnotesize\textbf{\textit{Entry Point Widget}}}
This widget is the simplest set as it provides the entry point only by a static yellow cylinder of radius \textit{r}= 1\,mm \textit{length}=3\,mm. The dentist proposed it due to missing rotational alignment information and feedback. It is a negative control condition that resembles the current "unassisted" method. Since it is known from the literature, it should be included to increase comparability \cite{tao2023comparison,ma2019augmented,lin2015novel}.

\paragraph{\footnotesize\textbf{\textit{Target Axis Widget}}}
    This widget is most abundant in related works\cite{tao2023comparison,dastan2022gestalt,ma2019augmented,wang2014augmented,katic2010knowledge}. It is a yellow cylinder of \textit{r}= 1\,mm, \textit{length}=120\,mm with a yellow color for easy identification. The intersection point of the axis with the gingiva provides spatial positioning like the entry point condition, while the axis represents the drill path in space. A centered red disc on top of the drill tool enables alignment with the target axis, similar to the work by Song et al. \cite{song2018endodontic}. Unlike the static Entry Point, it provides pseudo-static feedback of correct spatial positioning and serves as a baseline.

\subsubsection {Dynamic Widgets (DWs)}
Dynamic Widget (DW) \footnote{DW open source link: \url{https://github.com/Vr3xMelab/DW.git}.} provides real-time dynamic feedback of the spatial positional and rotational error, taking inspiration from the study of Dastan et al. \cite{dastan2022gestalt}, which uses the gestalt theory of perception \cite{lim2007interaction}. This approach mimics the behavior of analog measurement tools like mechanical calipers and photographic optical viewfinders.
The DW's basic principle is to provide visual feedback using graphical error visualization \cref{fig: Total DW}. The span error between tool and target is reflected by the mutual distance of forms (\cref{fig: dw non collimated,fig: dw COLLIMATED}), three positional "\textbf{V}"-shaped duos (one per axis component) (\cref{fig: DW V}) and two rotational "\textbf{(}"-shaped duos (one per 5DOF component) (\cref{fig: DW rot}). 
During co-design, we discussed the work of Dastan et al. \cite{dastan2022gestalt} and were inspired to implement our DW. We improved the following features:
\begin{itemize} 
     \item optimization of dynamic visibility areas' threshold parameters;
     \item applied nonlinear law for the displacement of related form duos;
     \item implementation of 3D forms instead of 2D textured icons;
    \item attached the DW closer to the drill tooltip to improve intuitiveness;
    \item the DW rotation form duos are related to the user's controller rotation instead of locked to the world reference system;
    \item applied custom shader materials to form duos to prevent visual occlusion by other visual objects in the scene.
\end{itemize}

\begin{figure} [t]
    \centering
    \includegraphics[width=\columnwidth]{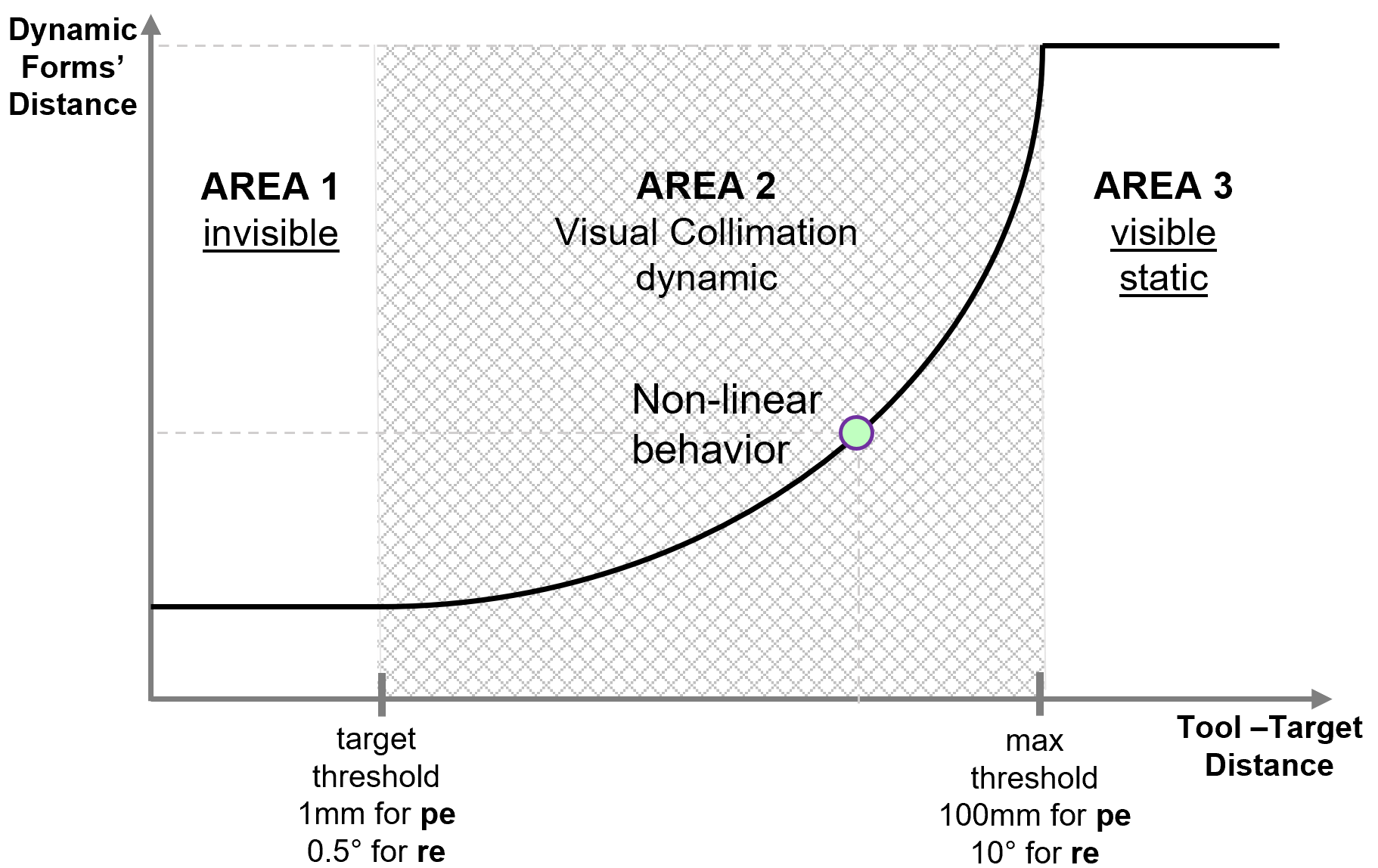}
    \caption{Implementation of three dynamic visibility areas: Area 1: Invisible forms, Area 2: Visible Collimation Area with dynamic and non-linear behavior, Area 3: Visible static forms.}
    \label{fig: Non-linear}
\end{figure}

\begin{figure}
\centering
    \includegraphics[width=\columnwidth]{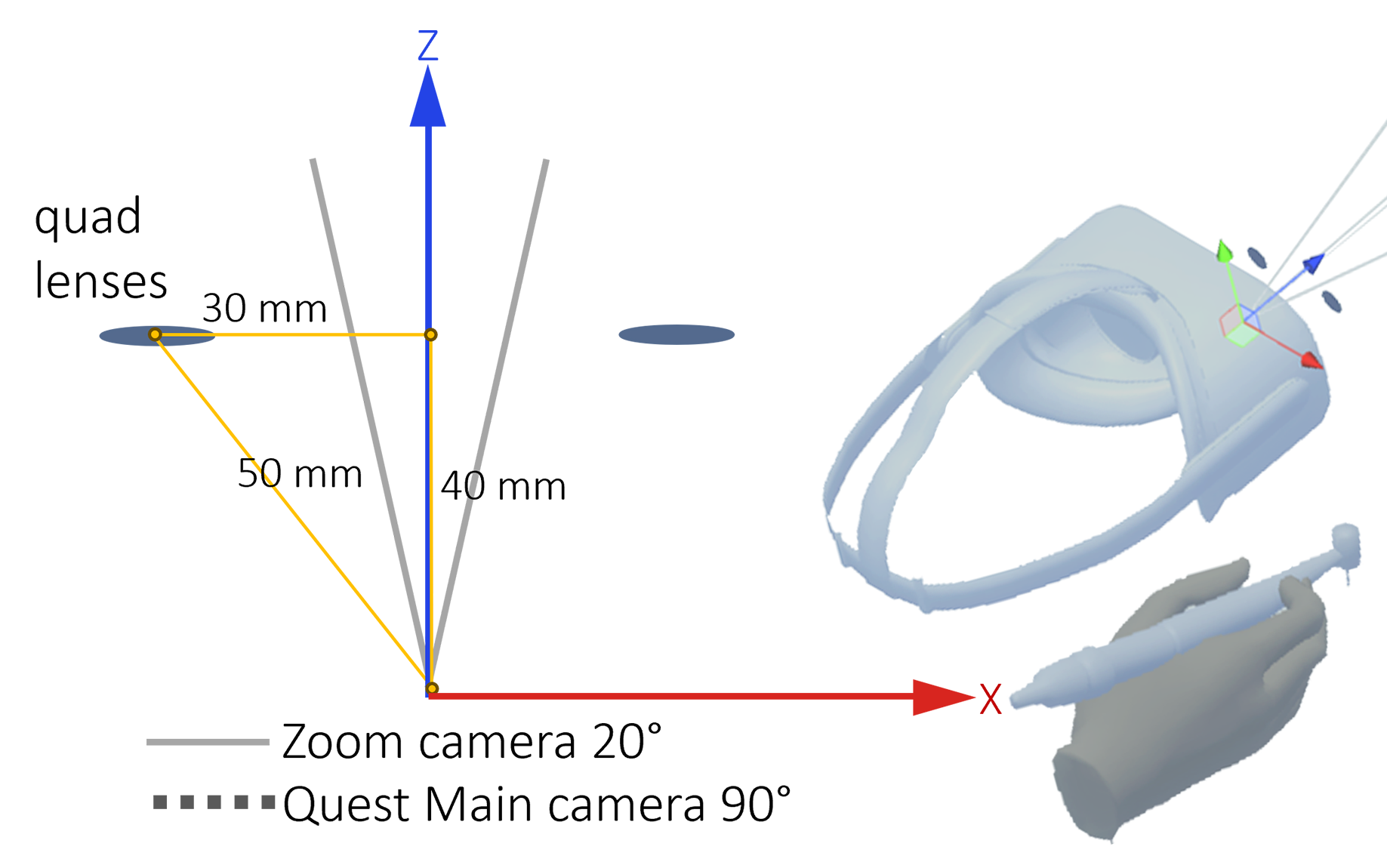}
    \caption{Implementation of the virtual magnifying loupe; top view of lenses' angles and positions (left), two quads attached to the headset(right).}
    \label{fig:loupe}
\end{figure}
\begin{figure} [ht]
    \centering
    \includegraphics[width=\columnwidth]{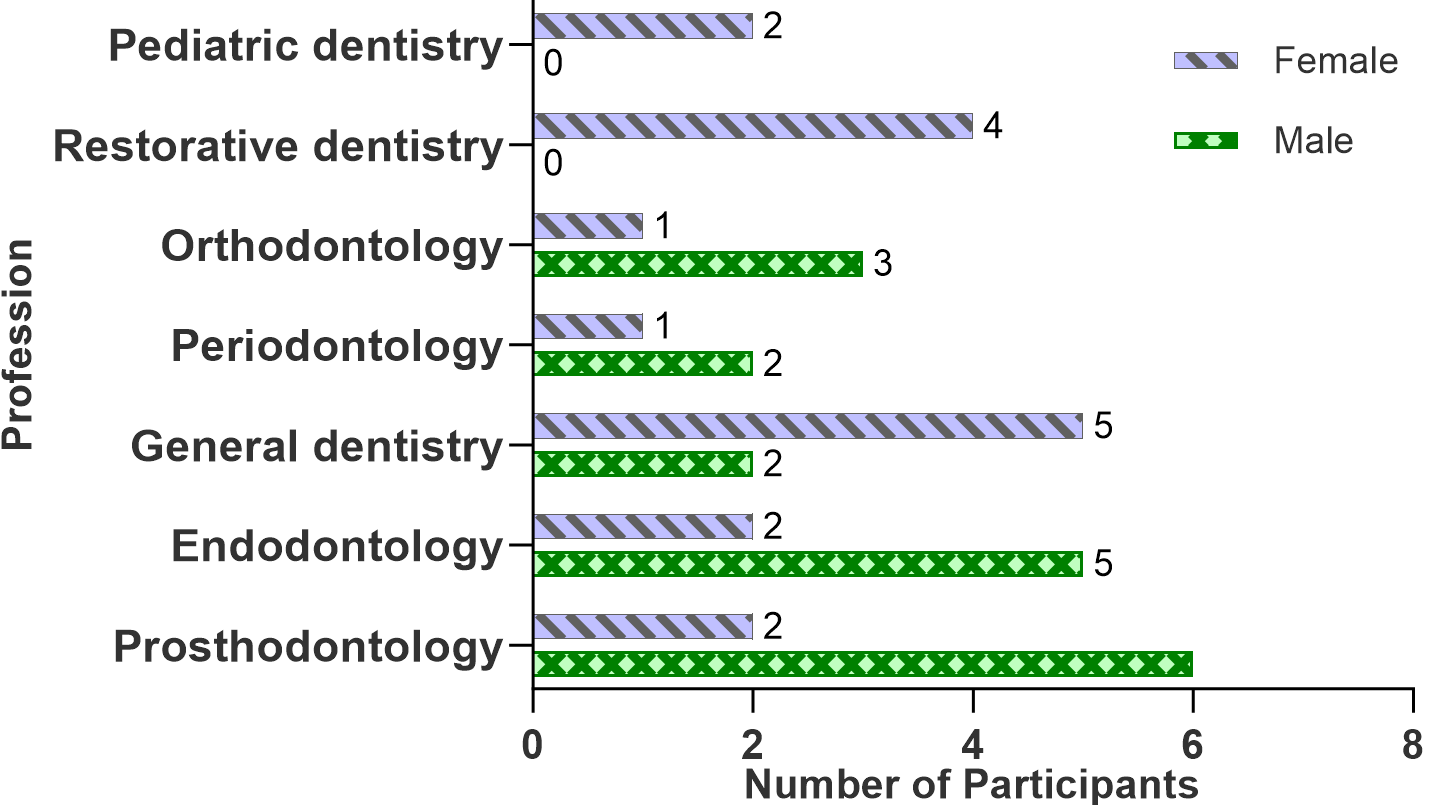}
    \caption{Gender/profession of dentists participated in the user study.}
      \label{fig: profession}
\end{figure}

\textbf{Dynamic visibility areas:} 
We defined the positional vectors as tool position (\textbf{top}) and target position (\textbf{tap}) and the rotations as tool rotation (\textbf{tor}) and target rotation (\textbf{tar}). We also split the spatial positional error into pure positional error (\textbf{pe}) and the rotational error (\textbf{re}) of the tool and target.
We modified the dynamic visibility behavior of the Dastan et al. \cite{dastan2022gestalt}. The main goal of this modification is to reduce visual overload. From our co-design, we assigned the following dynamic visibility areas (\cref{fig: Non-linear}):
\begin{itemize} [noitemsep]
    \item Area 1: Target threshold (tt) = 1\,mm for \textbf{pe} and 0.5\,° for \textbf{re}; these parameters set with dentists' feedback during co-design. The component is hidden when tt has been reached as the user needs no intervention.
    \item Area 2: Dynamic nonlinear behavior, duos are visualized. This reduces the amplification in the target vicinity and avoids overshooting fatigue (\cref{fig: Non-linear})
   \begin{equation}
  \label{eq:combined}
  \textbf{pe} = f_{\textbf{pe}}(\textbf{tap} - \textbf{top})^2 \quad \text{and} \quad \textbf{re} = f_{\textbf{re}}(\textbf{tor} \cdot \textbf{tar}^{-1})
\end{equation}
    \item Area 3: Pairs are visualized but static/frozen at a max threshold (mt) = 100\,mm for \textbf{pe} and 10\,° for \textbf{re}, co-designed with the dentists' feedback to avoid being too far away from sight.
\end{itemize}

\paragraph{\footnotesize\textbf{\textit{Dynamic Widget and Entry Point (DWEP)}}}
DWEP combines the DW with the Entry Point widget. This condition provides an almost "DW-only" scenario, where the entry point only indicates the target position.

\paragraph{\footnotesize\textbf{\textit{Dynamic Widget and Target Axis (DWTA)}}}
This condition combines DW and static Target Axis widgets. The mixed DWTA gives two references for the dentist to follow during tool positioning. DWTA evaluates the combined widgets' impact on performance and task load.

\section{Evaluation}
During co-design assessments, it was seen that DWTA might provide better objective and subjective parameters than the other three conditions. Additionally, we noticed that targets were reached faster using the Target Axis and Entry Point widgets; therefore, we hypothesized they would result in less cognitive effort.
A within-subject user study was conducted with dentists to objectively analyze our co-design results and compare MRDPWs: DWTA, DWEP, Target Axis, and Entry Point conditions. We evaluated the user performance, task load, and preference of MRDPWs implemented in co-design sessions. This section includes the details of the user study.

\begin{figure*} [t]
\centering
    \includegraphics[width=\textwidth]{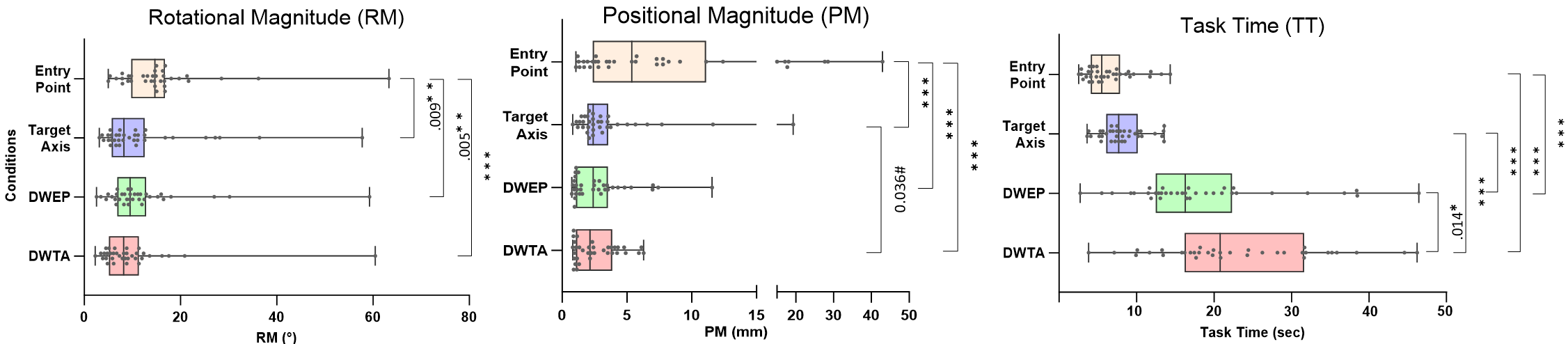}
    \caption{Conditions box plot comparisons of user study results (n=35, * = Friedman (pbonf***$\leq$0.001), \#= Wilcoxon (p\#$\leq$0.05) test): rotational magnitude RM and positional magnitude PM: \textbf{Entry Point} is the least accurate,  task time TT: Entry Point faster than \textbf{DWEP} and \textbf{DWTA}, \textbf{Target Axis} is faster than DWEP, DWTA and DWEP is faster than DWTA.}
    \label{fig:TIME-PM-RM}
    
\end{figure*}
\begin{figure}
    \centering
    \includegraphics[width=\columnwidth]{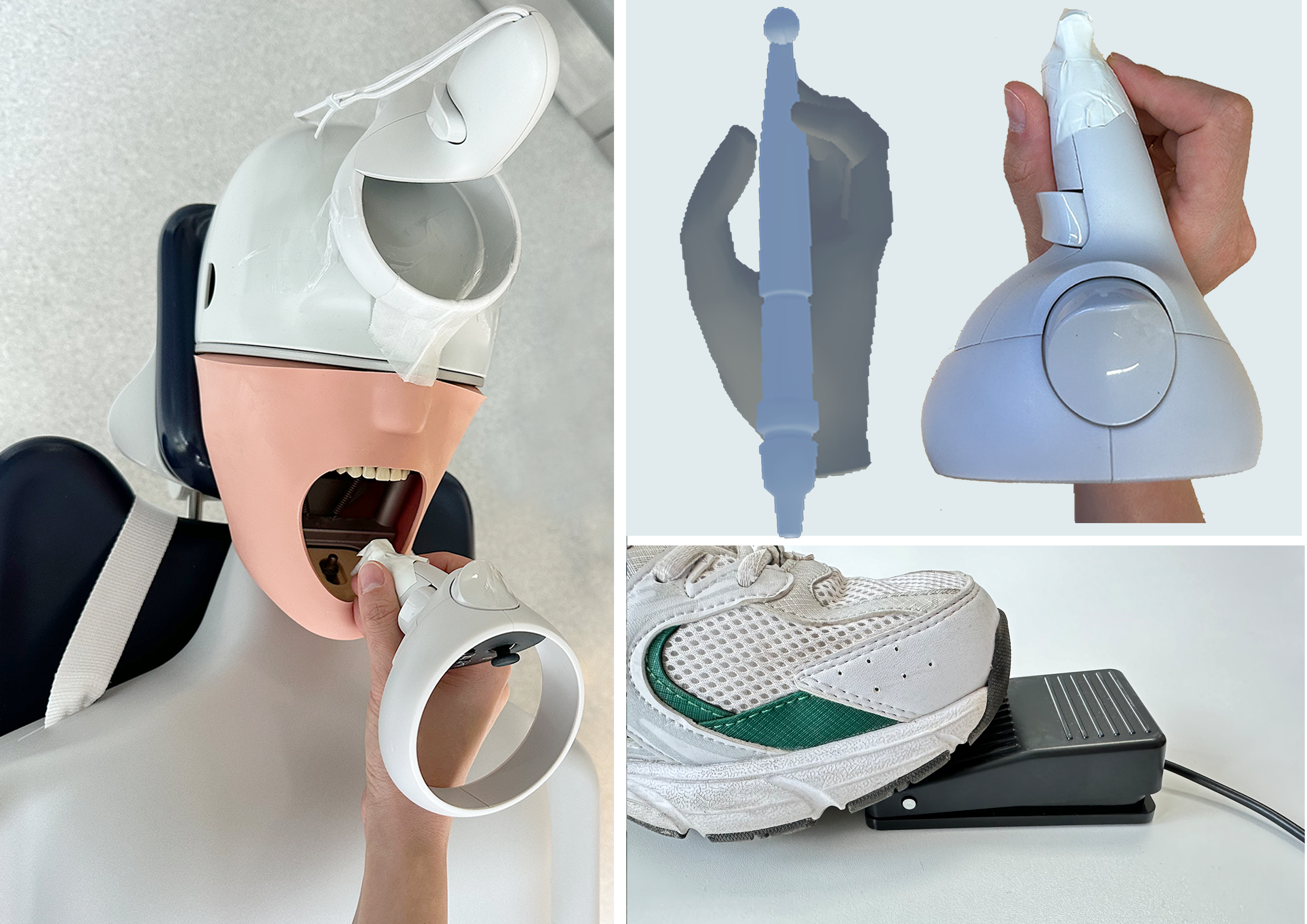}
    \caption{Experiment setup; Phantom model tracked with the left controller (left). The right hand is for tool handling (top-right), and the foot pedal is the input device (bottom-right).}
    \color{black}
    \label{fig: setup}
\end{figure}
\begin{figure} [ht]
    \centering
    \includegraphics[width=\columnwidth]{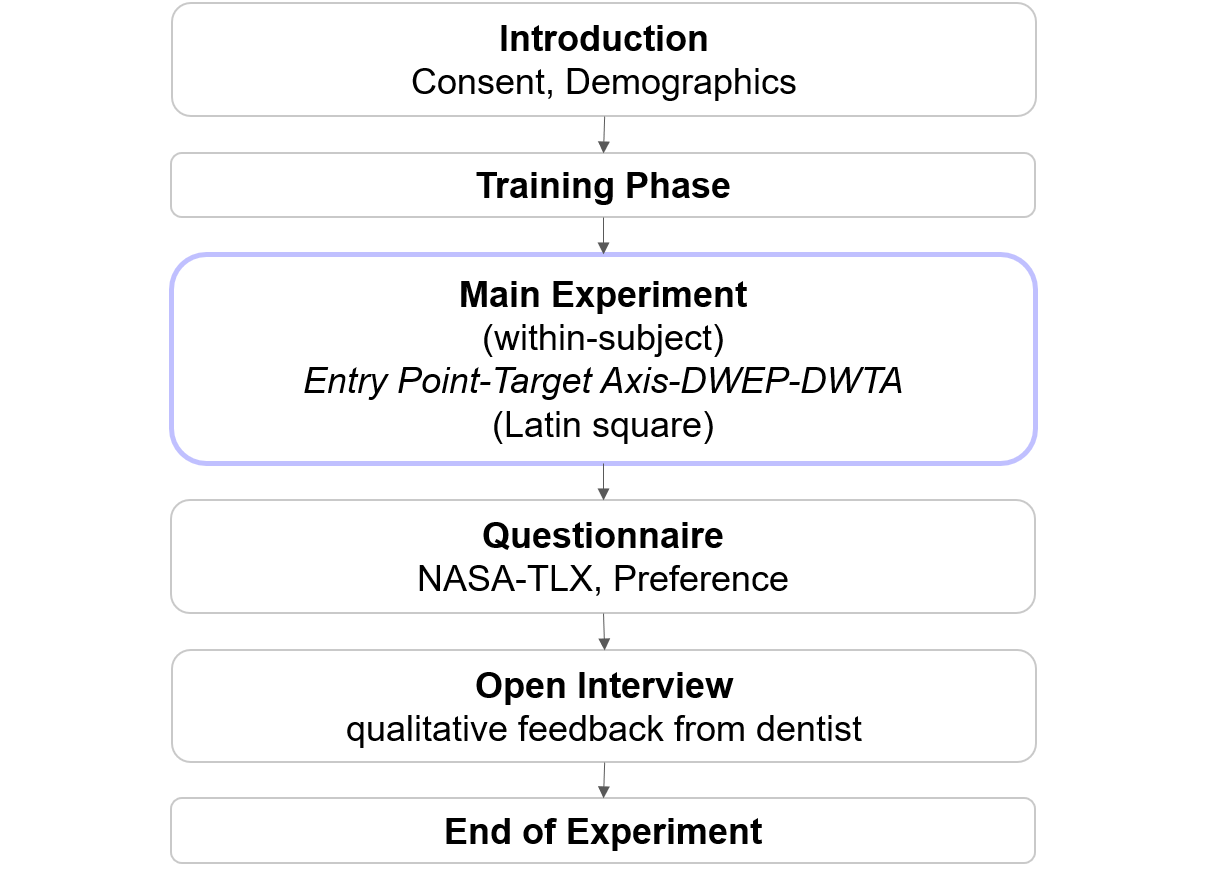}
    \caption{User Study protocol overview.}
    \label{fig: procedure}
\end{figure}
\subsection{Hypotheses}
We stated below the research hypotheses (H$_n$) following our co-design evaluation:

We set the following hypotheses for RQ$_2$ and RQ$_3$:
     \begin{itemize} [noitemsep]
    \item \textbf{H$_1$}: DWTA has better precision than other conditions.
    \item \textbf{H$_2a$}: DWEP is faster than the DWTA, 
    \item \textbf{H$_2b$}: DWTA has better precision than DWEP.
    \item \textbf{H$_3$}: Entry Point "non-assisted method" has the lowest task time.
    \item  \textbf{H$_4$}: Target Axis has less task load than DWs.
    \item  \textbf{H$_5$}: The added complexity of DW increases the task load.
     \end{itemize}
     
We set the following hypothesis for RQ$_4$:
 \begin{itemize}
    \item   \textbf{H$_6$}: There is a negative correlation between precision and time.
     \end{itemize}
     
We set the following hypothesis for RQ$_5$:
     \begin{itemize}
    \item   \textbf{H$_7$}: Younger generations and subjects with gaming experience have better precision using DWs.
     \end{itemize} 

\subsection{Participants}
Our study was approved by our institutional review board. Informed consent was obtained from each participant. Thirty-five voluntary dentists participated (18\Male,\;17\Female) aged 34.34 ± 9.9 years. They belong to various specialties: Eight in prosthodontology, seven in endodontology, seven in general dentistry, three in periodontology, four in orthodontology, four in restorative dentistry, and two in pediatric dentistry \cref{fig: profession}. This group doesn't include the first two dentists participating in the co-design phase.  We collected participants' ages, years of experience in dentistry, AR/VR experience, and gaming rates.

\begin{figure*} 
  \centering
    \includegraphics[width=\textwidth]{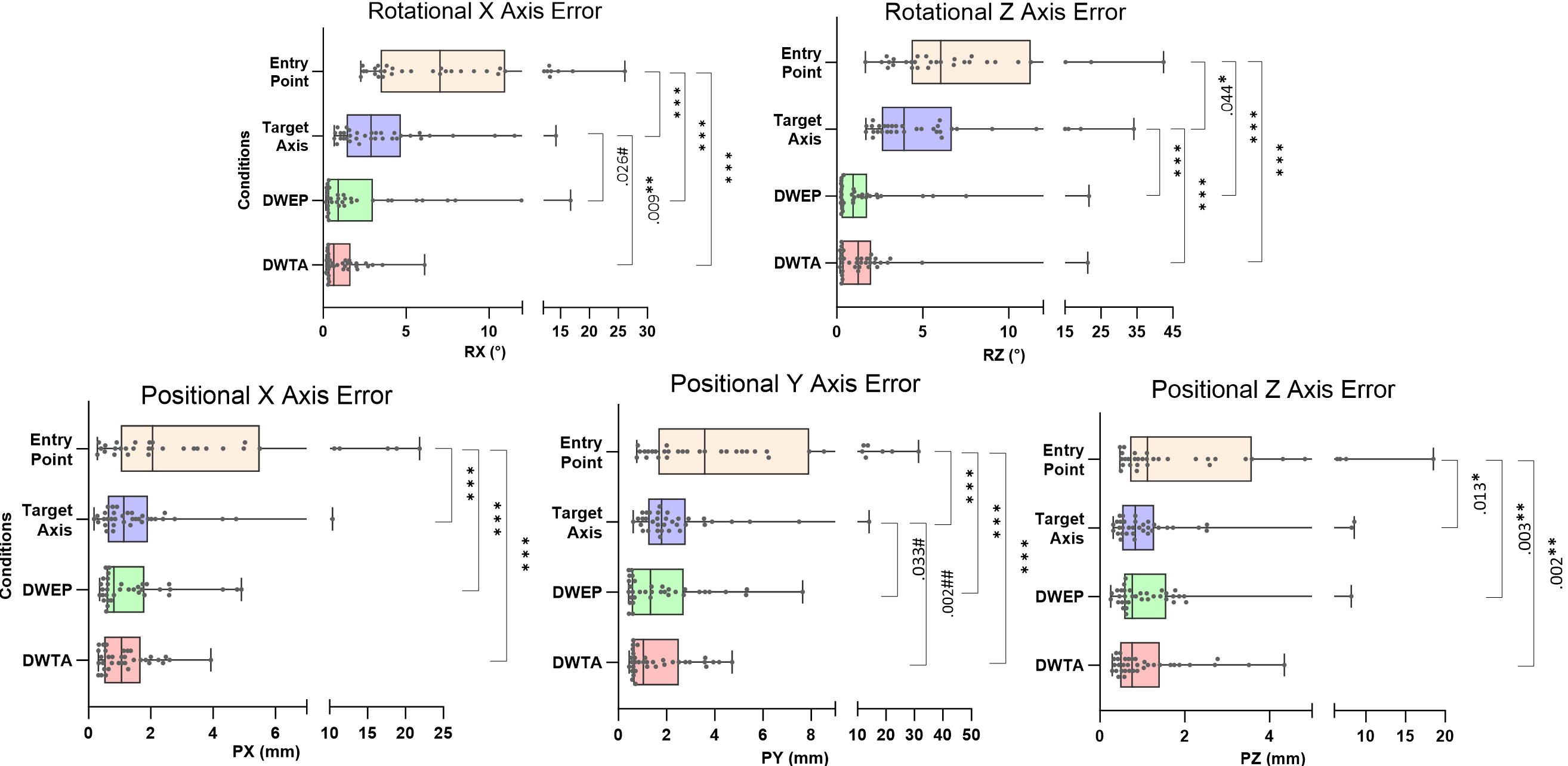}
        \caption{Conditions Box plot comparisons of user study results (n=35, * = Friedman (pbonf***$\leq$0.001), \#= Wilcoxon (p\#$\leq$0.05) test) for Rotational X and Z Axis errors (RZ and RX) and Positional X, Y, and Z errors (PX, PY, PZ): \textbf{Entry Point} is the least precise in  RX, RZ, PX, PY, PZ. \textbf{Target Axis} is less precise than \textbf{DWTA} in RX, RZ. Target Axis is less precise than \textbf{DWEP} in RZ.}
    \label{fig:rxrypmpxpz}
\end{figure*}

\subsection{Co-designed Setup}
We used an AR simulated in VR setup as supported by its validity in literature \cite{lee2009replication,harris2020framework,lee2010role}. As a result of co-design, we built a realistic setup, supported in a physical dental studio, to imitate the body pose and movements. The hardware comprised an Oculus Quest 2 (Meta, CA, USA) HMD, which was connected to a laptop PC with a USB-A cable, a Foot Switch USB-A foot pedal (Xiatiaosann, Guangzhou, China), and two Meta Quest 2 controllers (Meta Platforms). The right-hand controller was represented in VR by a virtual hand grasping a drill tool (\cref{fig: setup}).
To achieve realistic drill handling, users turned the right-hand controller and held it backward. A handmade drill tooltip was attached to the right-hand controller to mimic the drill tool.
The left controller tracked the virtual patient's head position on the real phantom to provide arm/hand rest reference to the participants (\cref{fig: setup}).

{\textbf{\textit{Dental Room and Phantom}}}
We used a dental room with a physical phantom as described in literature \cite{tao2023comparison,ma2019augmented, lin2015novel, wang2014augmented}. The dental unit's backrest angulation was adjusted to a horizontal position, and the physical phantom was affixed to the dental unit with belts (\cref{fig: teaser}). The dental stool was positioned on the right side of the phantom and parallel to the dental unit. The dentists were asked to sit on the stool to simulate the operative stance.
In VR, we designed a dental room arrangement with a static avatar patient and dental unit. All targets are positioned only on the virtual patient's mandible.

{\textbf{\textit{Foot pedal}}} 
We simulated the dentist's drill control using Unity 3D input action linked to the USB-based pedal (\cref{fig: setup}). Participants confirmed the tool positioning on each target by pressing the pedal. After each pedal hit, sound feedback was provided; afterward, the next target appeared in the mandible.

{\textbf{\textit{Virtual Loupe}}}
Dentists suggested a VR magnifying loupe during the co-design process, which is also supported by literature with positive outcomes in surgical tasks \cite{aldosari2021dental,branson2018using,wajngarten2021magnification,qian2020ar,martin2009head}. In each condition, the virtual loupe was attached to the headset camera for a focused task field of view to ensure better posture and minimize physical strain.
The VR loupe was implemented using two circle quads (r= 10\,mm), a zoom camera, and render-texture material in Unity. Each quad has a 30\,mm distance from its center point. Quads were attached to the main camera with a 50 cm offset and followed the head movement (\cref{fig:loupe}). Using post-processing in the main camera, we added a blur effect to enhance the scene's depth perception.

\subsection{Procedure and Drill Positioning Task}
We introduced the experiment research problem and the four widgets, with a neutral and unbiased description of their function, followed by the test procedure instructions (\cref{fig: procedure}).
Participants sat on the dental stool and wore the Oculus Quest 2 headset. The experiment commenced with a training scenario in which each condition was presented sequentially. Afterward, the main experiment started. A balanced Latin square was used to establish the order of the MRDPW conditions (\cref{fig: teaser}).

The user task involved positioning and orienting the drill to the indicated target as precisely as possible.  An active depth drilling task was not included. The foot pedal was pressed once the dentist was satisfied with the precise drill tool positioning. The pedal hit saved the error value entry and proceeded with the next target. Targets appeared singly in random order. Each condition involved 16 repetitive tool positioning tasks for each subject.
After completing all positioning tasks for one condition, the next scene was launched. Objective metrics were automatically saved. Following the main experiment, participants were tasked with completing a subjective questionnaire. Subsequently, we sought dentists' opinions on the specific condition/setup/widget through an open interview session.
The entire user study took approximately 40-45 minutes.

\subsection{Metrics}
We collected the following parameters during our user study:

\textbf{Positional Error:} The positional magnitude (PM) was calculated to assess the precision of distance error between the drill tool-tip and the entry point target. Separated metrics for the distance along the x, y, and z-axis (PX, PY, PZ) were measured in millimeters.

\textbf{Rotational Error:} The rotational magnitude (RM) was calculated to assess the precision of rotational error between the drill tooltip and entry point axes. Separate metrics for the delta angle along the x and z-axis (RX, RZ) were measured in degrees.

\textbf{Task Time:} The elapsed time between the single target display and foot pedal press, measured in seconds. 

\textbf{Task load:} After the experiment, participants compiled a NASA-TLX \cite{hart2006nasa} questionnaire (Likert 7 scale), including mental demand, physical demand, temporal demand, effort, and frustration.

\textbf{User Preference and Opinions:} 
Following the questionnaire, we conducted post-experiment open interviews with the participants to record their preferences and gather their opinions.

\subsection{Statistics}
Data was analyzed using JASP software (version 0.18.3.0, University of Amsterdam). Normality was assessed using the Shapiro-Wilk test, which revealed a non-Gaussian data distribution. First, descriptive data analysis was performed for RM, PM, and TT. Consequently, we employed the non-parametric Friedman test to assess statistical significance for paired samples alongside Kendall's W to measure the effect size. Bonferroni corrections were used for post hoc comparisons (pbonf$\leq$.05). We further conducted a Wilcoxon test to compare two independent samples that had a mean difference (md), whether their mean ranks statistically differ (p$\leq$.05). Only significant comparisons from the Wilcoxon test are reported in the results section. Additionally, we calculated Pearson's correlation coefficient (r) to measure the linear relationship between precision and time variables.  Significance was assumed at r-value$\leq$0.6:strong,$\leq$0.4:moderate,$\leq$0.2:weak,<0.2:very weak.

\section{Results}
\subsection{Objective Measures}
 All participants performed all tasks of precision-focused drill tool positioning on sequential targets for MRDPW conditions. In total, we collected 2240 samples (16 tasks x 4 conditions x 35 participants) for objective measures. The results are summarized below in a concise manner:
\begin{figure} [t]
    \centering
    \includegraphics[width=\columnwidth]{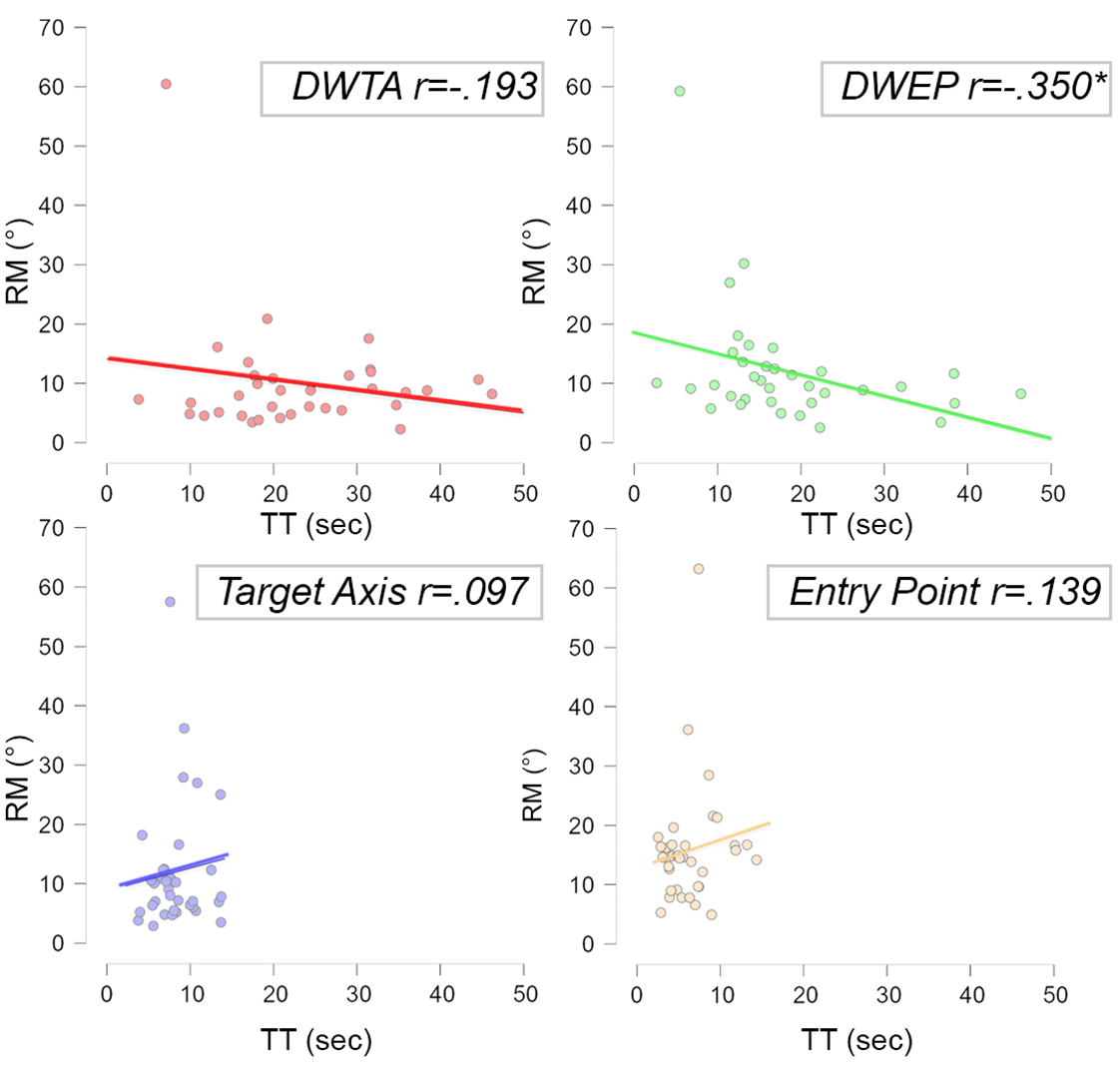}
    \caption{RM-TT Correlations: DWEP has a significant negative correlation; RM precision requires time. Correlation strengths (r) are presented (r-value$\leq$0.6:strong,$\leq$0.4:moderate,$\leq$0.2:weak,<0.2:very weak).}
    \label{fig:condition RM-TT}
\end{figure}

\begin{figure} [t]
    \centering
    \includegraphics[width=\columnwidth]{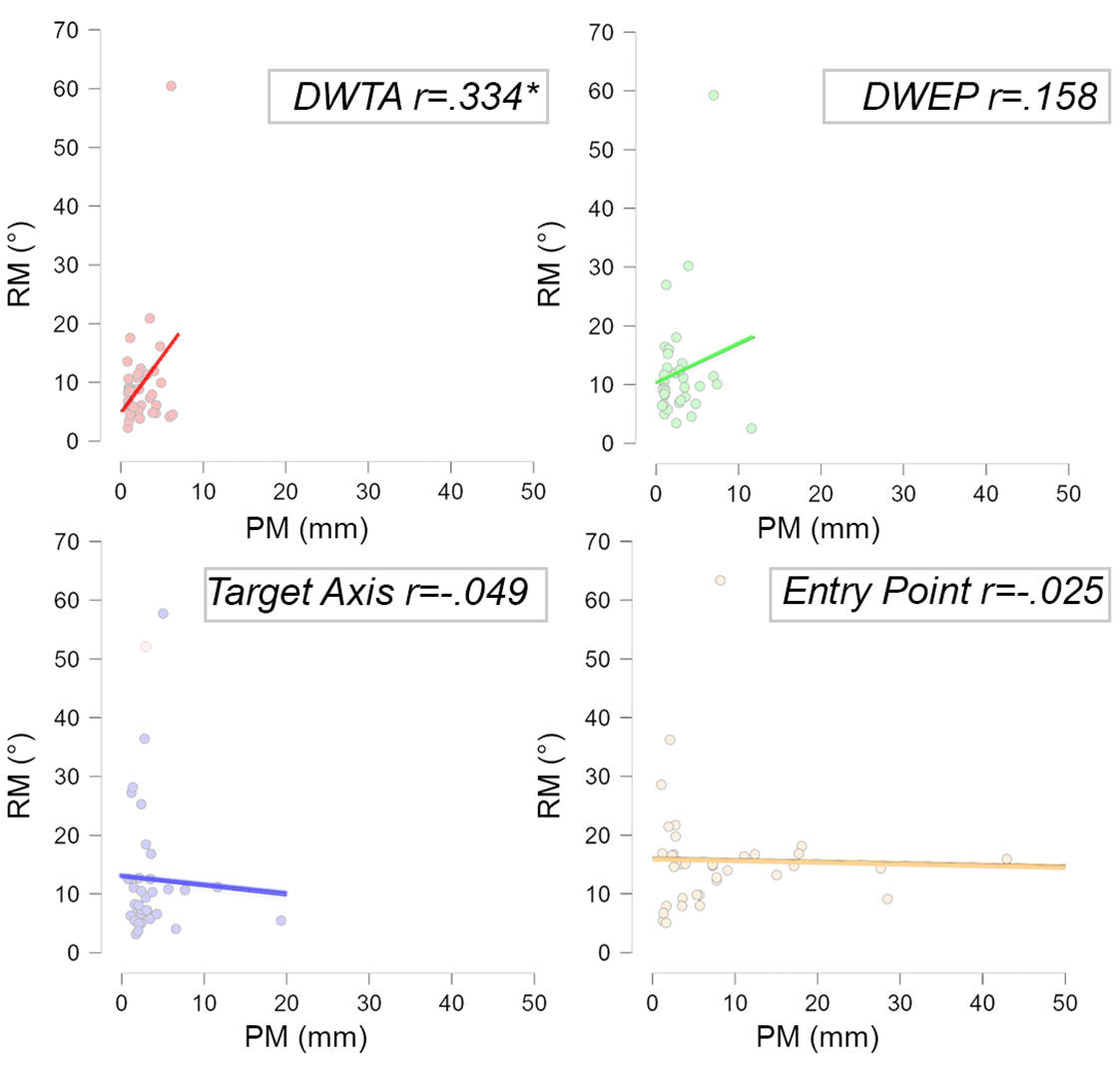}
    \caption{RM-PM Correlations: DWTA has a significant positive correlation pointing to better RM precision leading to better PM precision.}
    \label{fig:condition RM-PM}
\end{figure}

\paragraph{\footnotesize\textbf{\textit{Rotational Error: RM, RX, RZ}}}

RM: The descriptive statistics are DWTA (9.96\,° ± 9.72), DWEP (12.09\,° ± 10.02), Target Axis ( 12.29\,° ± 11.01), Entry Point (15.94\,° ± 10.32). The Friedman test revealed statistical significance for RM ($x^2$= 35.81, p<.001, Kendall's W= 0.34). Post hoc Dunn-Bonferroni corrections highlighted that Enry Point has the least RM precision (\cref{fig:TIME-PM-RM}).

RX: The Friedman test demonstrated significance in RX ($x^2$= 62.72, p<.001, Kendall's W= 0.59). Post hoc Dunn-Bonferroni corrections indicated that Entry Point is the least precise condition, and Target Axis is less precise than DWTA (\cref{fig:rxrypmpxpz}). Further, a Wilcoxon test revealed significance in DWEP<Target Axis (md =1.22, z=-2.22, p=.026*).

RZ: The Friedman test indicated significance for RZ ($x^2$= 70.26, p<.001, Kendall's W= 0.66). Post hoc Dunn-Bonferroni corrections demonstrated that Entry Point is the least precise condition and Target Axis is less precise than DWEP and DWTA (\cref{fig:rxrypmpxpz}).

\paragraph{\footnotesize\textbf{\textit{Positional Error: PM, PX, PY, PZ}}}
PM: The descriptive statistics are DWTA (2.58\,mm ± 1.69), DWEP (2.85\,mm ± 2.41), Target Axis (3.55\,mm ± 0.81) and Entry Point (8.43\,mm ± 1.06). The Friedman test yielded significant results for PM ($x^2$= 19.90, p<.001, Kendall's W= 0.19). Subsequent post hoc Dunn-Bonferroni corrections were applied. The results demonstrated that Entry Point is a less precise condition (\cref{fig:TIME-PM-RM}). While non-significant median values were observed, a Wilcoxon test revealed significance for DWTA < Target Axis comparisons (md= 0.96, z=2.09, and p=.036).

PX The Friedman test demonstrated statistical significance in PX ($x^2$ = 15.65, p=.001, Kendall's W= 0.15). Post hoc Dunn-Bonferroni corrections revealed that Entry Point is the least precise condition (\cref{fig:rxrypmpxpz}).

PY  The Friedman test indicated statistical significance for PY ($x^2$= 24.46, p<.001, Kendall's W= 0.23). Post hoc Dunn-Bonferroni corrections revealed that Entry Point is the least precise condition (\cref{fig:rxrypmpxpz}). Further, the Wilcoxon paired samples t-test showed DWEP<Target Axis (md= 0.86, z=2.12, p=.033*) and DWTA<Target Axis (md=0.25, z=2.99, p=.002**).

PZ: The Friedman test yielded statistical significance in PZ ($x^2$= 12.90, p=.005, Kendall's W= 0.12). Post hoc Dunn-Bonferroni corrections demonstrated that Entry Point is the least precise condition (\cref{fig:rxrypmpxpz}).

\paragraph{\footnotesize\textbf{\textit{Task Time}}}
The descriptive statistics for TT; Entry Point (6.39\,sec ± 3.02), Target Axis (8.17\,sec ± 2.78), DWEP (18.46\,sec ± 9.88), DWTA (23.03\,sec ± 10.32).  The Friedman test revealed statistical significance ($x^2$= 75.61, p<.001, Kendall's W =0.72). Post hoc Dunn-Bonferroni corrections applied results yielded significance for Entry Point < DWEP, DWTA (pbonf <.001***), Target Axis< DWEP, DWTA (pbonf<.001***) and DWEP < DWTA (pbonf=.014*) (\cref{fig:TIME-PM-RM}).

\subsection{Subjective Measures}
\paragraph{\footnotesize\textbf{\textit{Observations}}}
During the experiment, the observer noted the following events. Five participants exhibited excessive body movement in Entry Point and Target Axis conditions. One user asked for a larger virtual loupe. One experienced blurry vision, and another reported eye strain. A 61-year-old dentist experienced neck problems and took longer to complete the task.

\begin{figure}[hbt!]
    \centering
    \includegraphics[width=\columnwidth]{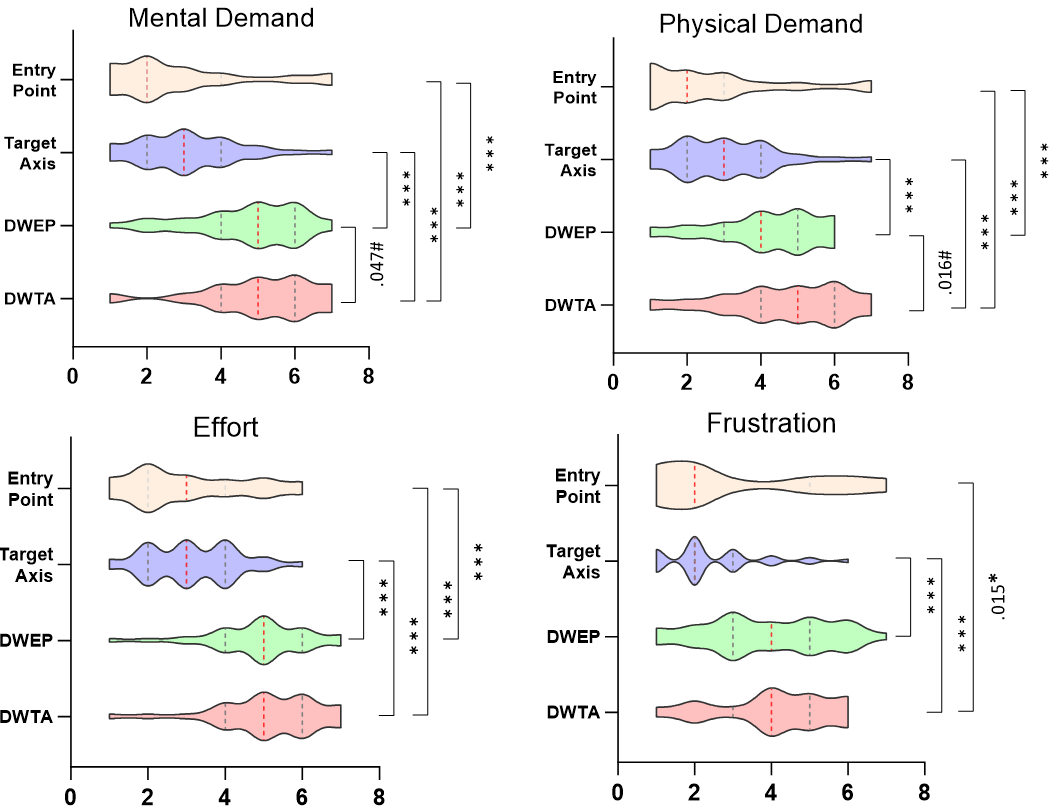}
    \caption{NASA-TLX results, (n=35, * = Friedman (pbonf***$\leq$0.001), \#= Wilcoxon (p\#$\leq$0.05) test), Entry Point and Target Axis have lower task load than DWEP and DWTA.}
    \label{fig: NASA}
\end{figure}

\paragraph{\footnotesize\textbf{\textit{NASA-TLX}}}
Participants compiled the questionnaire about the task load (\cref{fig: NASA}), age, background, year of experience, gaming skill rate, and preference.
On average, they had nine years of experience in dentistry 5\,$\bar{X}$  (min 1, max 36 ± 9\,IQR). They rated their gaming skills as 3.31 ± 2.01 / 7 and their knowledge of AR/VR applications as 2.7 ± 1.75 / 7 (having tried AR/VR apps at least once). All were right-handed, except two left-handed and one ambidextrous. Nevertheless, all could use the right hand and the right foot for the task. 

Mental Demand: Entry Point and Target Axis had lower mental demand than DWEP and DWTA (p<.001***). DWEP<DWTA (p=.047*) (\cref{fig: NASA}).

Physical Demand: Entry Point and Target Axis had lower physical demand than DWEP and DWTA (p<.001***). DWEP<DWTA (p=.016*)(\cref{fig: NASA}).

Effort: Entry Point and Target Axis required less effort than DWEP and DWTA (p<.001***) (\cref{fig: NASA}).

Frustration: Entry Point had less frustration than DWTA, and Target Axis had less frustration than DWEP or DWTA (p<.001***) (\cref{fig: NASA}).

\section{Discussion}

\begin{figure} [t]
    \centering
    \includegraphics[width=\columnwidth]{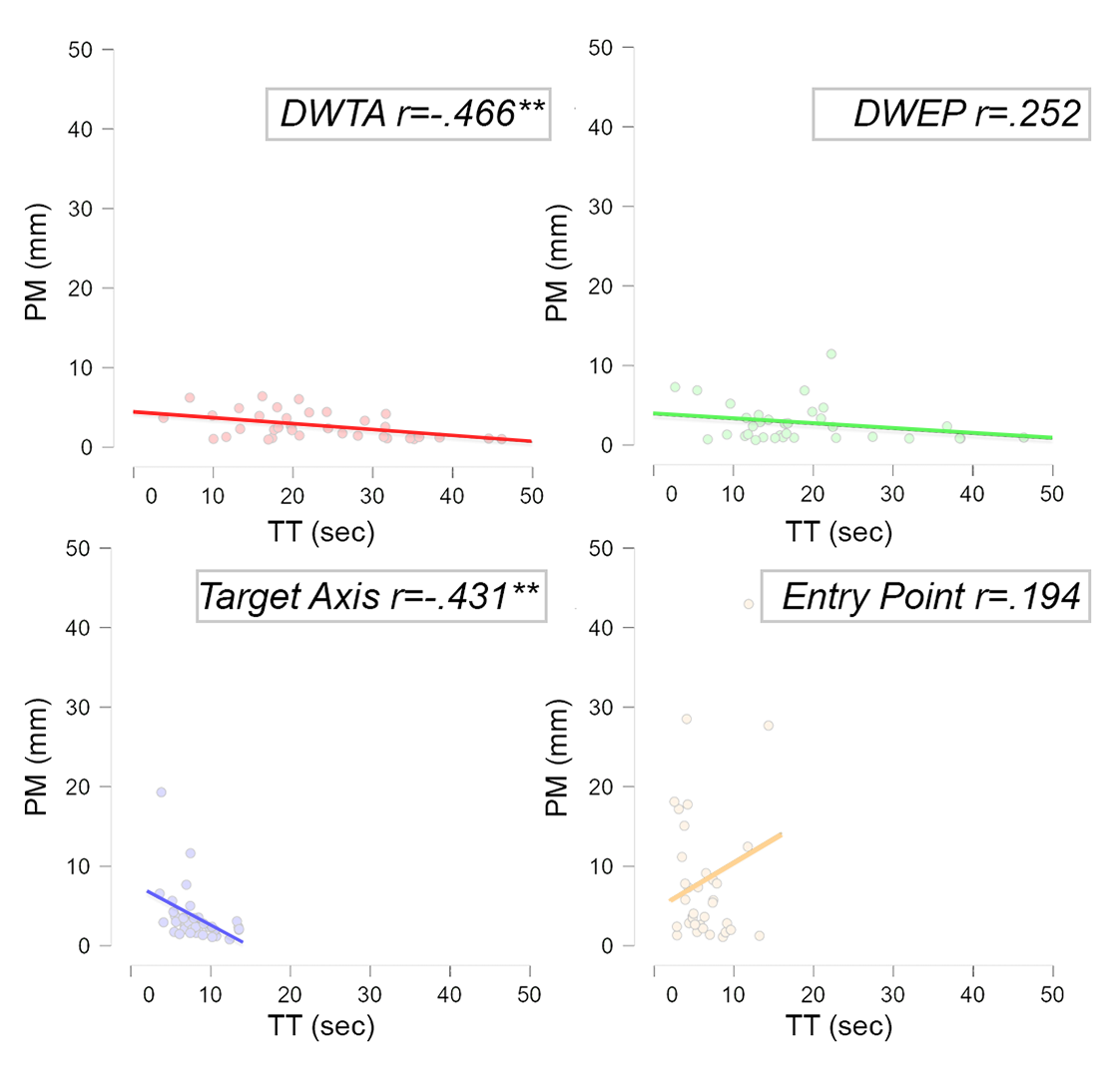}
    \caption{Correlations of PM-TT: DWTA and Target Axis demonstrate negative correlation, revealing that more time trade-off ensures better PM precision.}
    \label{fig: correlation PM-TT}
\end{figure}
\begin{figure} [t]
    \centering
    \includegraphics[width=\columnwidth]{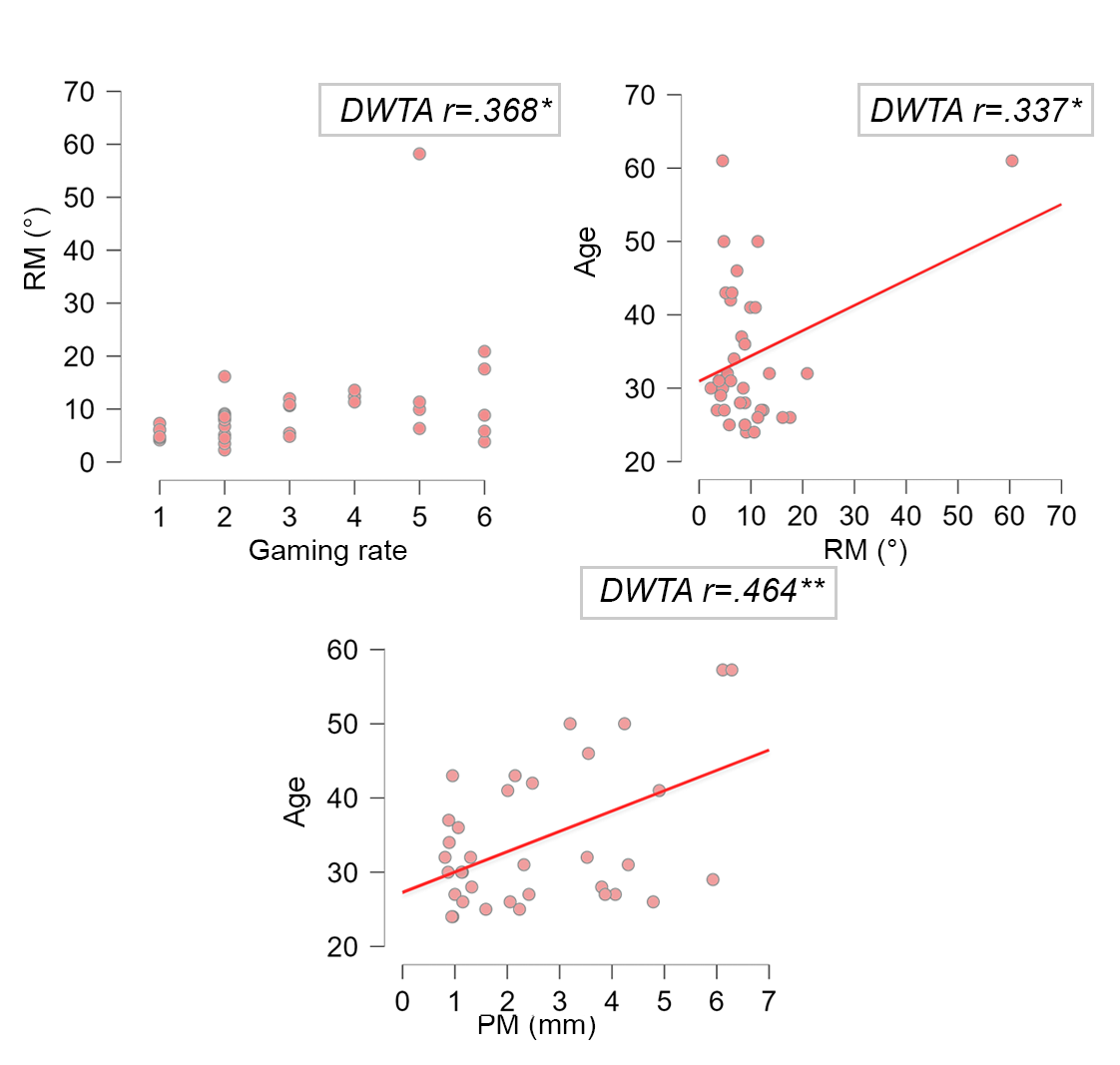}
    \caption{Positive DWTA correlations: Subjects with better gaming experience, younger generations have more precision.}
    \label{fig: DWTA correlation}
\end{figure}
 \begin{figure*} [t]
     \centering
    \includegraphics [width=\textwidth]{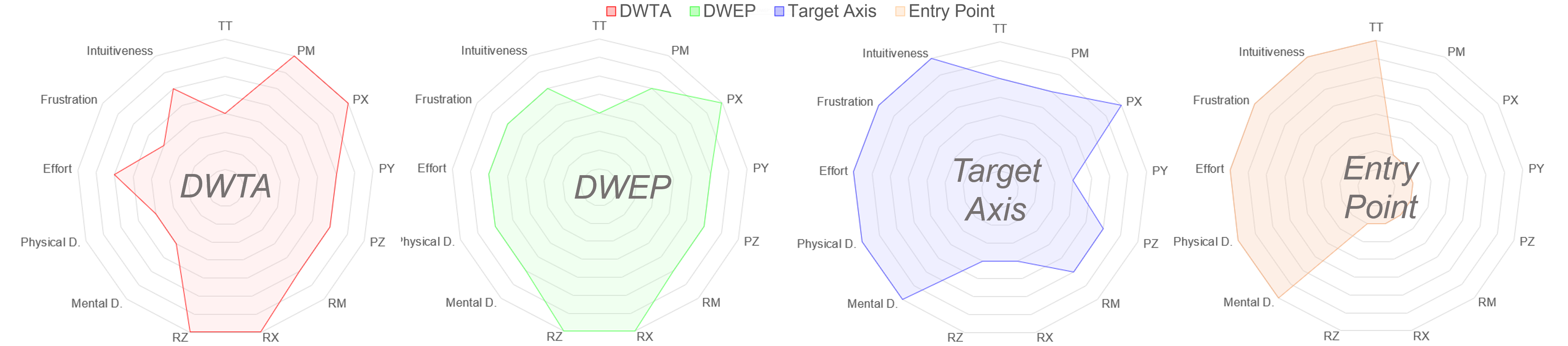}
    \caption{Radar-plot graphics for MRDPW conditions overview; Entry Point, Target Axis, DWEP, DWTA, and user preference written in the center of each plot(\textit{up}).}
    \label{fig:Radarplot}
\end{figure*}
\subsection{Objective Measures Analyses}

In addressing our research questions, we observed significant improvement of DWs over static Target Axis in positional and rotational precision. Both DWs outperform Target Axis in positional PX and rotational  RX and RZ. Furthermore, DWTA is more effective in PM than Target Axis, which supports our \textbf{H$_1$}.

On the other hand, DWs have a time trade-off comparing SWs. This outcome aligns with existing literature \cite{dastan2022gestalt} in which DWs are significantly more precise but slower than SWs. This is an expected and interesting finding as it highlights the value of DWs in providing better precision and the trade-off in execution time. Interestingly, \textbf{H$_2a$} is supported, and DWEP is faster than DWTA, indicating that more complex widget design influences task time. However, unlike our expectations, \textbf{H$_2b$} is not supported, and the two dynamic widgets did not differ statistically in positional and rotational precision.

Besides, our \textbf{H$_3$} is supported; the Entry Point is faster than the other conditions. However, although the widget shows the exact entry point, the Entry Point is the least precise condition in PM and for each axis component (PX, PY, and PZ). Target Axis provides better positional precision than Entry Point, indicating that Entry Point requires additional information to indicate the target position. Entry Point is also the least precise in rotational precision (RM, RX, and RZ), which is evident since the widget doesn't include rotational hints. 

During the co-design phase, dentists stressed the importance of keeping positional and rotational error low as key factors compared to task time. We see that the quantitative results are similar and in line with the literature regarding the advantages of MR in assisted tool positioning, \cite{tao2024comparison,ma2019augmented,dastan2022gestalt}. Additionally, we executed correlations between time-positional and rotational precisions to investigate MRDPWs' performances further.
The DWTA (r=-.466**) and Target Axis (r=-.431**) are related by the significant negative correlation between PM and TT (\textbf{H$_6$}), proving the effect of spending more time results in better positional precision (\cref{fig: correlation PM-TT}).

DWTA (r=.334*) shows a significant positive correlation between RM and PM, demonstrating that attention to the widget can bring better precision in position and rotation (\cref{fig:condition RM-PM}).

DWEP (r=-.350*) demonstrates a significant negative correlation between RM and TT, which supports our \textbf{H$_6$}, proving that more time is acceptable to reach rotational precision (\cref{fig:condition RM-TT}). These results highlight that the 3D dynamic widgets provide better PM and RM precision at the price of a time trade-off.

Our \textbf{H$_7$} is also supported; when comparing age and precision for DWTA, junior participants performed with less PM (r=.464**) and RM (r=.337*) precision. This result can be explained by the minor familiarity of senior dentists with utilizing complex widgets.
In support of this, subjects with gaming experience perform better for the DWTA  (r=.368*) (\cref{fig: DWTA correlation}), also indicating their ability to understand the most complex widget of the set.

\subsection{Subjective Measures Analyses}
\paragraph{\footnotesize\textit{\textbf{NASA-TLX}}}
As expected, Entry Point and Target Axis, being the simplest to understand and the fastest to execute, resulted in lower Mental, Physical, and Effort demand scores than dynamic DWTA and DWEP, which partially supports \textbf{H$_4$}. The frustration rate was lower for Target Axis than DWTA-DWEP, while Entry Point was DWTA only. These results indicate that Target Axis is relevant in literature \cite{dastan2022gestalt} as an immediate and simple to understand widget. Also, our \textbf{H$_5$} is partially supported, and simpler DWEP has less mental and physical demand than DWTA.
These results differ from the previous study by \cite{dastan2022gestalt}, in which our DWs yielded more mental demand and frustration than SWs.

\paragraph{\footnotesize\textit{\textbf{Preference}}}
Target Axis singularly is the preferred with 16\,/35 dentists. It is curious to notice that this preference comes from junior and senior dentists (Age of experience 3\,$\bar{X}$, min 1-max 36, 13.25\,IQR). Despite this preference, observers noted that the Target Axis and Entry Point required more body movement and position changes during the task. Furthermore, participants mentioned that \textit{P$_1$$_0$:}"\textit{Target Axis is straightforward however it is challenging to align rotations precisely}" and \textit{P$_1$$_8$:}"\textit{Target Axis was perceived as irritating and made orientation difficult.}"

DWTA was preferred from 12\,/35 dentists of all seniority levels (Age of experience 6.25\,$\bar{X}$, min 1-max 34, 5.75\,IQR). This choice supports the idea that having direct references to the point and axis of Target Axis is a key graphical cue for this widget. Besides, 7\,/35 participants (Age of experience 4\,$\bar{X}$, min 1-max 17, 12\,IQR) preferred DWEP. Participants stated about DWs; \textit{P$_7$:} "\textit{With training the DW could potentially improve usability}," and \textit{P$_1$$_4$:} "\textit{DW could enhance precision; training with the DW is the optimal choice}." 
One user praised the  DW's spatial support: \textit{P$_1$$_6$:}"\textit{The frontal axis was challenging to comprehend; the DW is a valid solution.}". Another participant without prior experience with AR/VR declared that \textit{P$_1$$_7$:}"\textit{DW is challenging, Target Axis is demanding to perceive orientation}."

Interestingly, Entry Point has no preferences. It is remarkable how real usage showcased the limitations. \textit{P$_7$:} "\textit{I felt insecure about the precision}" and \textit{P$_2$$_5$:}"\textit{Entry Point is perceived as uncertain and more demanding for precision}." This indicates that dentists want additional navigation during the tool positioning and teaches us how MR interface design must confront previous procedures to be accepted by end users.

\paragraph{\footnotesize\textit{\textbf{Overview}}}
Although the Target Axis is the most preferred, DWs aggregated preference (12+7) is superior to SWs (19 vs. 16). This result may indicate the potential of DWs, but the current limits of proposed implementations in managing effort vs performance tradeoff.
To enhance the comparisons of MRDPWs across multiple parameters, we employed radar plot graphs as illustrated in \cref{fig:Radarplot}. Each condition has been assigned a rating (+1 point) based on the results, significantly outperforming the other conditions. This graph presents a nuanced perspective on the strengths and weaknesses of each widget compared to others.

\subsection{Limitations and Future Work}
In this study, we focused on the co-design of MRDPWs. The efficacy of our designs was evaluated solely by the tool positioning task. Several other tasks relevant to the surgical workflow, such as active drilling or precision maintenance, have not yet been included and could influence the subjective and objective performance of the MRDPWs.
Moreover, in our setup, the mandibular model was absent in the phantom mouth, resulting in less haptic feedback than in a realistic setting. Additionally, the realistic patient movement could not be included since the study was conducted in \textit{in vitro}. The AR simulation in VR was implemented for two primary reasons: Firstly, our study focuses on the visual and dynamic feedback that can be effectively simulated and evaluated within a VR environment. Secondly, the technological limitations of current AR devices (tracking, precision, and latency) may influence the repeatability of the results. However, widgets' performance may be confirmed in real AR setups in future studies. Another limitation is the increased frustration of the presented dynamic widgets. We want to address it in future works with a direct design indicating the path to follow. And exploring different DW referential position configurations (attached to the target or displayed as a screen fixed).

\subsection{Takeaways}
The DW capability to improve dentist’s performances in realistic settings is a key result of this work but also sheds light on the interface trade-off between precision (position and rotation) and task load and its impact on user preferences. Furthermore, analyses reveal interesting correlations between user demographics (such as age and gaming experience) and familiarity with the proposed complex and dynamic interfaces. This aspect opens to the crucial role of training for dynamic and innovative widgets design DWs. Finally, we provide the DW scripts and all graphical assets with an open-source license to foster future research in replication, evaluation, and improvement of next-gen interfaces.

\section{Conclusions}
We carried out the MR dynamic widgets' co-design process for drill positioning involving two expert dentists and three MR  experts. We compared two dynamic variants to two static MRDPWs.
The multidisciplinary process resulted in benefits for dynamic widgets' positional and rotational precision and a trade-off regarding mental and physical effort and frustration. We can conclude that the design presented is optimal since the best DWs had more preference. A more direct affordance supported by cognitive perception theory must be investigated further. The value of this research demonstrates that DW's designs is documented modular and easy to reuse. In future scenarios, DWs can be easily applied to other medical or industrial scenarios \cite{burstrom2021augmented,waelkens2016surgical,eom2022neurolens,joda2019augmented,kaplan2021effects}, including manufacturing, assembly maintenance, with clear benefits in terms of safety, efficiency and better quality of life for the workers.

\section*{Supplemental Material}
Full version quantitative  results, video presentation and DW's open source link provided as supplemental materials \url{https://github.com/Vr3xMelab/DW.git}.

\acknowledgments{%
Support from the Italian Ministry of Education, University and Research (MUR) under the program ‘‘Departments of Excellence’’ (L.232/2016). The authors express their gratitude to the dentists at the Department of Conservative Dentistry and Periodontology, LMU Klinikum Hospital, Munich, for their contribution of the user study.}

\bibliographystyle{abbrv-doi-hyperref}

\bibliography{template}

\begin{thebibliography}{10}

\bibitem{aldosari2021dental}
M.~A. Aldosari.
\newblock Dental magnification loupes: an update of the evidence.
\newblock {\em The Journal of Contemporary Dental Practice}, 22(3):310--315, 2021. \href{https://doi.org/10.5005/jp-journals-10024-3057}
{doi: {{%
10\hspace{.1pt}\discretionary{.}{%
}{.}\hspace{.4pt}5005\discretionary{/}{%
}{/}jp\discretionary{%
}{-}{-}journals\discretionary{%
}{-}{-}10024\discretionary{%
}{-}{-}3057}}}


\bibitem{bertollo2011drilling}
N.~Bertollo and W.~R. Walsh.
\newblock Drilling of bone: practicality, limitations and complications associated with surgical drill-bits.
\newblock {\em Biomechanics in applications}, 4:53--83, 2011. \href{https://doi.org/10.5772/20931}
{doi: {{%
10\hspace{.1pt}\discretionary{.}{%
}{.}\hspace{.4pt}5772\discretionary{/}{%
}{/}20931}}}


\bibitem{besanccon2021state}
L.~Besan{\c{c}}on, A.~Ynnerman, D.~F. Keefe, L.~Yu, and T.~Isenberg.
\newblock The state of the art of spatial interfaces for 3d visualization.
\newblock In {\em Computer Graphics Forum}, vol.~40, pp. 293--326. Wiley Online Library, 2021. \href{https://doi.org/10.1111/cgf.14189}
{doi: {{%
10\hspace{.1pt}\discretionary{.}{%
}{.}\hspace{.4pt}1111\discretionary{/}{%
}{/}cgf\hspace{.1pt}\discretionary{.}{%
}{.}\hspace{.4pt}14189}}}


\bibitem{bird2021generative}
M.~Bird, M.~McGillion, E.~Chambers, J.~Dix, C.~Fajardo, M.~Gilmour, K.~Levesque, A.~Lim, S.~Mierdel, C.~Ouellette, et~al.
\newblock A generative co-design framework for healthcare innovation: development and application of an end-user engagement framework.
\newblock {\em Research involvement and engagement}, 7:1--12, 2021. \href{https://doi.org/10.1186/s40900-021-00252-7}
{doi: {{%
10\hspace{.1pt}\discretionary{.}{%
}{.}\hspace{.4pt}1186\discretionary{/}{%
}{/}s40900\discretionary{%
}{-}{-}021\discretionary{%
}{-}{-}00252\discretionary{%
}{-}{-}7}}}


\bibitem{block2017implant}
M.~S. Block, R.~W. Emery, D.~R. Cullum, and A.~Sheikh.
\newblock Implant placement is more accurate using dynamic navigation.
\newblock {\em Journal of Oral and Maxillofacial Surgery}, 75(7):1377--1386, 2017. \href{https://doi.org/10.1016/j.joms.2017.02.026}
{doi: {{%
10\hspace{.1pt}\discretionary{.}{%
}{.}\hspace{.4pt}1016\discretionary{/}{%
}{/}j\hspace{.1pt}\discretionary{.}{%
}{.}\hspace{.4pt}joms\hspace{.1pt}\discretionary{.}{%
}{.}\hspace{.4pt}2017\hspace{.1pt}\discretionary{.}{%
}{.}\hspace{.4pt}02\hspace{.1pt}\discretionary{.}{%
}{.}\hspace{.4pt}026}}}


\bibitem{branson2018using}
B.~Branson, R.~Abnos, M.~Simmer-Beck, G.~King, and S.~Siddicky.
\newblock Using motion capture technology to measure the effects of magnification loupes on dental operator posture: a pilot study.
\newblock {\em Work}, 59(1):131--139, 2018. \href{https://doi.org/10.3233/WOR-172681}
{doi: {{%
10\hspace{.1pt}\discretionary{.}{%
}{.}\hspace{.4pt}3233\discretionary{/}{%
}{/}WOR\discretionary{%
}{-}{-}172681}}}


\bibitem{burstrom2021augmented}
G.~Burstr{\"o}m, O.~Persson, E.~Edstr{\"o}m, and A.~Elmi-Terander.
\newblock Augmented reality navigation in spine surgery: a systematic review.
\newblock {\em Acta Neurochirurgica}, 163:843--852, 2021. \href{https://doi.org/10.1007/s00701-021-04708-3}
{doi: {{%
10\hspace{.1pt}\discretionary{.}{%
}{.}\hspace{.4pt}1007\discretionary{/}{%
}{/}s00701\discretionary{%
}{-}{-}021\discretionary{%
}{-}{-}04708\discretionary{%
}{-}{-}3}}}


\bibitem{busciantella2024research}
D.~Busciantella-Ricci and S.~Scataglini.
\newblock Research through co-design.
\newblock {\em Design Science}, 10:e3, 2024. \href{https://doi.org/10.1017/dsj.2023.35}
{doi: {{%
10\hspace{.1pt}\discretionary{.}{%
}{.}\hspace{.4pt}1017\discretionary{/}{%
}{/}dsj\hspace{.1pt}\discretionary{.}{%
}{.}\hspace{.4pt}2023\hspace{.1pt}\discretionary{.}{%
}{.}\hspace{.4pt}35}}}


\bibitem{cassetta2020there}
M.~Cassetta, F.~Altieri, M.~Giansanti, M.~Bellardini, G.~Brandetti, and L.~Piccoli.
\newblock Is there a learning curve in static computer-assisted implant surgery? a prospective clinical study.
\newblock {\em International journal of oral and maxillofacial surgery}, 49(10):1335--1342, 2020. \href{https://doi.org/10.1016/j.ijom.2020.03.007}
{doi: {{%
10\hspace{.1pt}\discretionary{.}{%
}{.}\hspace{.4pt}1016\discretionary{/}{%
}{/}j\hspace{.1pt}\discretionary{.}{%
}{.}\hspace{.4pt}ijom\hspace{.1pt}\discretionary{.}{%
}{.}\hspace{.4pt}2020\hspace{.1pt}\discretionary{.}{%
}{.}\hspace{.4pt}03\hspace{.1pt}\discretionary{.}{%
}{.}\hspace{.4pt}007}}}


\bibitem{cassetta2017much}
M.~Cassetta and M.~Bellardini.
\newblock How much does experience in guided implant surgery play a role in accuracy? a randomized controlled pilot study.
\newblock {\em International Journal of Oral and Maxillofacial Surgery}, 46(7):922--930, 2017. \href{https://doi.org/10.1016/j.ijom.2017.03.010}
{doi: {{%
10\hspace{.1pt}\discretionary{.}{%
}{.}\hspace{.4pt}1016\discretionary{/}{%
}{/}j\hspace{.1pt}\discretionary{.}{%
}{.}\hspace{.4pt}ijom\hspace{.1pt}\discretionary{.}{%
}{.}\hspace{.4pt}2017\hspace{.1pt}\discretionary{.}{%
}{.}\hspace{.4pt}03\hspace{.1pt}\discretionary{.}{%
}{.}\hspace{.4pt}010}}}


\bibitem{chackartchi2022reducing}
T.~Chackartchi, G.~E. Romanos, L.~Parkanyi, F.~Schwarz, and A.~Sculean.
\newblock Reducing errors in guided implant surgery to optimize treatment outcomes.
\newblock {\em Periodontology 2000}, 88(1):64--72, 2022. \href{https://doi.org/10.1111/prd.12411}
{doi: {{%
10\hspace{.1pt}\discretionary{.}{%
}{.}\hspace{.4pt}1111\discretionary{/}{%
}{/}prd\hspace{.1pt}\discretionary{.}{%
}{.}\hspace{.4pt}12411}}}


\bibitem{conner1992three}
B.~D. Conner, S.~S. Snibbe, K.~P. Herndon, D.~C. Robbins, R.~C. Zeleznik, and A.~Van~Dam.
\newblock Three-dimensional widgets.
\newblock In {\em Proceedings of the 1992 symposium on Interactive 3D graphics}, pp. 183--188, 1992. \href{https://doi.org/10.1145/147156.147199}
{doi: {{%
10\hspace{.1pt}\discretionary{.}{%
}{.}\hspace{.4pt}1145\discretionary{/}{%
}{/}147156\hspace{.1pt}\discretionary{.}{%
}{.}\hspace{.4pt}147199}}}


\bibitem{dastan2022gestalt}
M.~Dastan, A.~E. Uva, and M.~Fiorentino.
\newblock Gestalt driven augmented collimator widget for precise 5 dof dental drill tool positioning in 3d space.
\newblock In {\em 2022 IEEE International Symposium on Mixed and Augmented Reality (ISMAR)}, pp. 187--195. IEEE, 2022. \href{https://doi.org/10.1109/ISMAR55827.2022.00033}
{doi: {{%
10\hspace{.1pt}\discretionary{.}{%
}{.}\hspace{.4pt}1109\discretionary{/}{%
}{/}ISMAR55827\hspace{.1pt}\discretionary{.}{%
}{.}\hspace{.4pt}2022\hspace{.1pt}\discretionary{.}{%
}{.}\hspace{.4pt}00033}}}


\bibitem{dure2021first}
M.~Dur{\'e}, F.~Berlinghoff, M.~Kollmuss, R.~Hickel, and K.~C. Huth.
\newblock First comparison of a new dynamic navigation system and surgical guides for implantology: an in vitro study.
\newblock {\em International Journal of Computerized Dentistry}, 24(1), 2021.

\bibitem{el2019survey}
F.~El~Jamiy and R.~Marsh.
\newblock Survey on depth perception in head mounted displays: distance estimation in virtual reality, augmented reality, and mixed reality.
\newblock {\em IET Image Processing}, 13(5):707--712, 2019. \href{https://doi.org/10.1049/iet-ipr.2018.5920}
{doi: {{%
10\hspace{.1pt}\discretionary{.}{%
}{.}\hspace{.4pt}1049\discretionary{/}{%
}{/}iet\discretionary{%
}{-}{-}ipr\hspace{.1pt}\discretionary{.}{%
}{.}\hspace{.4pt}2018\hspace{.1pt}\discretionary{.}{%
}{.}\hspace{.4pt}5920}}}


\bibitem{el2020can}
H.~El-Jarn and G.~Southern.
\newblock Can co-creation in extended reality technologies facilitate the design process?
\newblock {\em Journal of Work-Applied Management}, 12(2):191--205, 2020. \href{https://doi.org/10.1108/JWAM-04-2020-0022}
{doi: {{%
10\hspace{.1pt}\discretionary{.}{%
}{.}\hspace{.4pt}1108\discretionary{/}{%
}{/}JWAM\discretionary{%
}{-}{-}04\discretionary{%
}{-}{-}2020\discretionary{%
}{-}{-}0022}}}


\bibitem{elani2018trends}
H.~Elani, J.~Starr, J.~Da~Silva, and G.~Gallucci.
\newblock Trends in dental implant use in the us, 1999--2016, and projections to 2026.
\newblock {\em Journal of dental research}, 97(13):1424--1430, 2018. \href{https://doi.org/10.1177/0022034518792567}
{doi: {{%
10\hspace{.1pt}\discretionary{.}{%
}{.}\hspace{.4pt}1177\discretionary{/}{%
}{/}0022034518792567}}}


\bibitem{eom2022neurolens}
S.~Eom, D.~Sykes, S.~Rahimpour, and M.~Gorlatova.
\newblock Neurolens: augmented reality-based contextual guidance through surgical tool tracking in neurosurgery.
\newblock In {\em 2022 IEEE International Symposium on Mixed and Augmented Reality (ISMAR)}, pp. 355--364. IEEE, 2022. \href{https://doi.org/10.1109/ISMAR55827.2022.00051}
{doi: {{%
10\hspace{.1pt}\discretionary{.}{%
}{.}\hspace{.4pt}1109\discretionary{/}{%
}{/}ISMAR55827\hspace{.1pt}\discretionary{.}{%
}{.}\hspace{.4pt}2022\hspace{.1pt}\discretionary{.}{%
}{.}\hspace{.4pt}00051}}}


\bibitem{fahim2022augmented}
S.~Fahim, A.~Maqsood, G.~Das, N.~Ahmed, S.~Saquib, A.~Lal, A.~A.~G. Khan, and M.~K. Alam.
\newblock Augmented reality and virtual reality in dentistry: highlights from the current research.
\newblock {\em Applied Sciences}, 12(8):3719, 2022. \href{https://doi.org/10.3390/app12083719}
{doi: {{%
10\hspace{.1pt}\discretionary{.}{%
}{.}\hspace{.4pt}3390\discretionary{/}{%
}{/}app12083719}}}


\bibitem{farronato2019current}
M.~Farronato, C.~Maspero, V.~Lanteri, A.~Fama, F.~Ferrati, A.~Pettenuzzo, and D.~Farronato.
\newblock Current state of the art in the use of augmented reality in dentistry: A systematic review of the literature.
\newblock {\em BMC Oral Health}, 19(1):1--15, 2019. \href{https://doi.org/10.1186/s12903-019-0808-3}
{doi: {{%
10\hspace{.1pt}\discretionary{.}{%
}{.}\hspace{.4pt}1186\discretionary{/}{%
}{/}s12903\discretionary{%
}{-}{-}019\discretionary{%
}{-}{-}0808\discretionary{%
}{-}{-}3}}}


\bibitem{farronato2023novel}
M.~Farronato, A.~Torres, M.~S. Pedano, and R.~Jacobs.
\newblock Novel method for augmented reality guided endodontics: An in vitro study.
\newblock {\em Journal of Dentistry}, 132:104476, 2023. \href{https://doi.org/10.1016/j.jdent.2023.104476}
{doi: {{%
10\hspace{.1pt}\discretionary{.}{%
}{.}\hspace{.4pt}1016\discretionary{/}{%
}{/}j\hspace{.1pt}\discretionary{.}{%
}{.}\hspace{.4pt}jdent\hspace{.1pt}\discretionary{.}{%
}{.}\hspace{.4pt}2023\hspace{.1pt}\discretionary{.}{%
}{.}\hspace{.4pt}104476}}}


\bibitem{freudenthal2011collaborative}
A.~Freudenthal, T.~St{\"u}deli, P.~Lamata, and E.~Samset.
\newblock Collaborative co-design of emerging multi-technologies for surgery.
\newblock {\em Journal of biomedical informatics}, 44(2):198--215, 2011. \href{https://doi.org/10.1016/j.jbi.2010.11.006}
{doi: {{%
10\hspace{.1pt}\discretionary{.}{%
}{.}\hspace{.4pt}1016\discretionary{/}{%
}{/}j\hspace{.1pt}\discretionary{.}{%
}{.}\hspace{.4pt}jbi\hspace{.1pt}\discretionary{.}{%
}{.}\hspace{.4pt}2010\hspace{.1pt}\discretionary{.}{%
}{.}\hspace{.4pt}11\hspace{.1pt}\discretionary{.}{%
}{.}\hspace{.4pt}006}}}


\bibitem{galindo2017influence}
P.~Galindo-Moreno, M.~Padial-Molina, P.~Nilsson, P.~King, N.~Worsaae, A.~Schramm, and C.~Maiorana.
\newblock The influence of the distance between narrow implants and the adjacent teeth on marginal bone levels.
\newblock {\em Clinical oral implants research}, 28(6):704--712, 2017. \href{https://doi.org/10.1111/clr.12867}
{doi: {{%
10\hspace{.1pt}\discretionary{.}{%
}{.}\hspace{.4pt}1111\discretionary{/}{%
}{/}clr\hspace{.1pt}\discretionary{.}{%
}{.}\hspace{.4pt}12867}}}


\bibitem{greenstein2015nerve}
G.~Greenstein, J.~R. Carpentieri, and J.~Cavallaro.
\newblock Nerve damage related to implant dentistry: incidence, diagnosis, and management.
\newblock {\em Compend Contin Educ Dent}, 36(9):652--9, 2015.

\bibitem{harris2020framework}
D.~J. Harris, J.~M. Bird, P.~A. Smart, M.~R. Wilson, and S.~J. Vine.
\newblock A framework for the testing and validation of simulated environments in experimentation and training.
\newblock {\em Frontiers in Psychology}, 11:605, 2020. \href{https://doi.org/10.3389/fpsyg.2020.00605}
{doi: {{%
10\hspace{.1pt}\discretionary{.}{%
}{.}\hspace{.4pt}3389\discretionary{/}{%
}{/}fpsyg\hspace{.1pt}\discretionary{.}{%
}{.}\hspace{.4pt}2020\hspace{.1pt}\discretionary{.}{%
}{.}\hspace{.4pt}00605}}}


\bibitem{harrison2022implementing}
R.~Harrison, E.~Ni~She, and D.~Debono.
\newblock Implementing and evaluating co-designed change in health.
\newblock {\em Journal of the Royal Society of Medicine}, 115(2):48--51, 2022. \href{https://doi.org/10.1177/01410768211070206}
{doi: {{%
10\hspace{.1pt}\discretionary{.}{%
}{.}\hspace{.4pt}1177\discretionary{/}{%
}{/}01410768211070206}}}


\bibitem{hart2006nasa}
S.~G. Hart.
\newblock Nasa-task load index (nasa-tlx); 20 years later.
\newblock In {\em Proceedings of the human factors and ergonomics society annual meeting}, vol.~50, pp. 904--908. Sage publications Sage CA: Los Angeles, CA, 2006. \href{https://doi.org/10.1177/154193120605000909}
{doi: {{%
10\hspace{.1pt}\discretionary{.}{%
}{.}\hspace{.4pt}1177\discretionary{/}{%
}{/}154193120605000909}}}


\bibitem{henderson2011augmented}
S.~J. Henderson and S.~K. Feiner.
\newblock Augmented reality in the psychomotor phase of a procedural task.
\newblock In {\em 2011 10th IEEE international symposium on mixed and augmented reality}, pp. 191--200. IEEE, 2011. \href{https://doi.org/10.1109/ISMAR.2011.6092386}
{doi: {{%
10\hspace{.1pt}\discretionary{.}{%
}{.}\hspace{.4pt}1109\discretionary{/}{%
}{/}ISMAR\hspace{.1pt}\discretionary{.}{%
}{.}\hspace{.4pt}2011\hspace{.1pt}\discretionary{.}{%
}{.}\hspace{.4pt}6092386}}}


\bibitem{joda2019augmented}
T.~Joda, G.~Gallucci, D.~Wismeijer, and N.~U. Zitzmann.
\newblock Augmented and virtual reality in dental medicine: A systematic review.
\newblock {\em Computers in biology and medicine}, 108:93--100, 2019. \href{https://doi.org/10.1016/j.compbiomed.2019.03.012}
{doi: {{%
10\hspace{.1pt}\discretionary{.}{%
}{.}\hspace{.4pt}1016\discretionary{/}{%
}{/}j\hspace{.1pt}\discretionary{.}{%
}{.}\hspace{.4pt}compbiomed\hspace{.1pt}\discretionary{.}{%
}{.}\hspace{.4pt}2019\hspace{.1pt}\discretionary{.}{%
}{.}\hspace{.4pt}03\hspace{.1pt}\discretionary{.}{%
}{.}\hspace{.4pt}012}}}


\bibitem{jung2008systematic}
R.~E. Jung, B.~E. Pjetursson, R.~Glauser, A.~Zembic, M.~Zwahlen, and N.~P. Lang.
\newblock A systematic review of the 5-year survival and complication rates of implant-supported single crowns.
\newblock {\em Clinical oral implants research}, 19(2):119--130, 2008. \href{https://doi.org/10.1111/j.1600-0501.2007.01453.x}
{doi: {{%
10\hspace{.1pt}\discretionary{.}{%
}{.}\hspace{.4pt}1111\discretionary{/}{%
}{/}j\hspace{.1pt}\discretionary{.}{%
}{.}\hspace{.4pt}1600\discretionary{%
}{-}{-}0501\hspace{.1pt}\discretionary{.}{%
}{.}\hspace{.4pt}2007\hspace{.1pt}\discretionary{.}{%
}{.}\hspace{.4pt}01453\hspace{.1pt}\discretionary{.}{%
}{.}\hspace{.4pt}x}}}


\bibitem{kaplan2021effects}
A.~D. Kaplan, J.~Cruit, M.~Endsley, S.~M. Beers, B.~D. Sawyer, and P.~A. Hancock.
\newblock The effects of virtual reality, augmented reality, and mixed reality as training enhancement methods: A meta-analysis.
\newblock {\em Human factors}, 63(4):706--726, 2021. \href{https://doi.org/10.1177/0018720820904229}
{doi: {{%
10\hspace{.1pt}\discretionary{.}{%
}{.}\hspace{.4pt}1177\discretionary{/}{%
}{/}0018720820904229}}}


\bibitem{katic2010knowledge}
D.~Kati{\'c}, G.~Sudra, S.~Speidel, G.~Castrillon-Oberndorfer, G.~Eggers, and R.~Dillmann.
\newblock Knowledge-based situation interpretation for context-aware augmented reality in dental implant surgery.
\newblock In {\em International Workshop on Medical Imaging and Virtual Reality}, pp. 531--540. Springer, 2010. \href{https://doi.org/10.1007/978-3-642-15699-1_56}
{doi: {{%
10\hspace{.1pt}\discretionary{.}{%
}{.}\hspace{.4pt}1007\discretionary{/}{%
}{/}978\discretionary{%
}{-}{-}3\discretionary{%
}{-}{-}642\discretionary{%
}{-}{-}15699\discretionary{%
}{-}{-}1\_56}}}


\bibitem{kivovics2022accuracy}
M.~Kivovics, A.~Tak{\'a}cs, D.~P{\'e}nzes, O.~N{\'e}meth, and E.~Mijiritsky.
\newblock Accuracy of dental implant placement using augmented reality-based navigation, static computer assisted implant surgery, and the free-hand method: an in vitro study.
\newblock {\em Journal of Dentistry}, 119:104070, 2022. \href{https://doi.org/10.1016/j.jdent.2022.104070}
{doi: {{%
10\hspace{.1pt}\discretionary{.}{%
}{.}\hspace{.4pt}1016\discretionary{/}{%
}{/}j\hspace{.1pt}\discretionary{.}{%
}{.}\hspace{.4pt}jdent\hspace{.1pt}\discretionary{.}{%
}{.}\hspace{.4pt}2022\hspace{.1pt}\discretionary{.}{%
}{.}\hspace{.4pt}104070}}}


\bibitem{lee2010role}
C.~Lee, S.~Bonebrake, D.~A. Bowman, and T.~H{\"o}llerer.
\newblock The role of latency in the validity of ar simulation.
\newblock In {\em 2010 IEEE Virtual Reality Conference (VR)}, pp. 11--18. IEEE, 2010. \href{https://doi.org/10.1109/VR.2010.5444820}
{doi: {{%
10\hspace{.1pt}\discretionary{.}{%
}{.}\hspace{.4pt}1109\discretionary{/}{%
}{/}VR\hspace{.1pt}\discretionary{.}{%
}{.}\hspace{.4pt}2010\hspace{.1pt}\discretionary{.}{%
}{.}\hspace{.4pt}5444820}}}


\bibitem{lee2009replication}
C.~Lee, S.~Bonebrake, T.~Hollerer, and D.~A. Bowman.
\newblock A replication study testing the validity of ar simulation in vr for controlled experiments.
\newblock In {\em 2009 8th IEEE International Symposium on Mixed and Augmented Reality}, pp. 203--204. IEEE, 2009. \href{https://doi.org/10.1109/ISMAR.2009.5336464}
{doi: {{%
10\hspace{.1pt}\discretionary{.}{%
}{.}\hspace{.4pt}1109\discretionary{/}{%
}{/}ISMAR\hspace{.1pt}\discretionary{.}{%
}{.}\hspace{.4pt}2009\hspace{.1pt}\discretionary{.}{%
}{.}\hspace{.4pt}5336464}}}


\bibitem{lim2007interaction}
Y.-k. Lim, E.~Stolterman, H.~Jung, and J.~Donaldson.
\newblock Interaction gestalt and the design of aesthetic interactions.
\newblock In {\em Proceedings of the 2007 conference on Designing pleasurable products and interfaces}, pp. 239--254, 2007. \href{https://doi.org/10.1145/1314161.1314183}
{doi: {{%
10\hspace{.1pt}\discretionary{.}{%
}{.}\hspace{.4pt}1145\discretionary{/}{%
}{/}1314161\hspace{.1pt}\discretionary{.}{%
}{.}\hspace{.4pt}1314183}}}


\bibitem{lin2015novel}
Y.-K. Lin, H.-T. Yau, I.-C. Wang, C.~Zheng, and K.-H. Chung.
\newblock A novel dental implant guided surgery based on integration of surgical template and augmented reality.
\newblock {\em Clinical implant dentistry and related research}, 17(3):543--553, 2015. \href{https://doi.org/10.1111/cid.12119}
{doi: {{%
10\hspace{.1pt}\discretionary{.}{%
}{.}\hspace{.4pt}1111\discretionary{/}{%
}{/}cid\hspace{.1pt}\discretionary{.}{%
}{.}\hspace{.4pt}12119}}}


\bibitem{ma2023visualization}
L.~Ma, T.~Huang, J.~Wang, and H.~Liao.
\newblock Visualization, registration and tracking techniques for augmented reality guided surgery: a review.
\newblock {\em Physics in Medicine \& Biology}, 68(4):04TR02, 2023. \href{https://doi.org/10.1088/1361-6560/acaf23}
{doi: {{%
10\hspace{.1pt}\discretionary{.}{%
}{.}\hspace{.4pt}1088\discretionary{/}{%
}{/}1361\discretionary{%
}{-}{-}6560\discretionary{/}{%
}{/}acaf23}}}


\bibitem{ma2019augmented}
L.~Ma, W.~Jiang, B.~Zhang, X.~Qu, G.~Ning, X.~Zhang, and H.~Liao.
\newblock Augmented reality surgical navigation with accurate cbct-patient registration for dental implant placement.
\newblock {\em Medical \& biological engineering \& computing}, 57:47--57, 2019. \href{https://doi.org/10.1007/s11517-018-1861-9}
{doi: {{%
10\hspace{.1pt}\discretionary{.}{%
}{.}\hspace{.4pt}1007\discretionary{/}{%
}{/}s11517\discretionary{%
}{-}{-}018\discretionary{%
}{-}{-}1861\discretionary{%
}{-}{-}9}}}


\bibitem{markovic2024considerations}
J.~Markovic, J.~F. Pe{\~n}a-Cardelles, I.~Pedrinaci, A.~Hamilton, G.~O. Gallucci, and A.~Lanis.
\newblock Considerations for predictable outcomes in static computer-aided implant surgery in the esthetic zone.
\newblock {\em Journal of Esthetic and Restorative Dentistry}, 36(1):207--219, 2024. \href{https://doi.org/10.1111/jerd.13171}
{doi: {{%
10\hspace{.1pt}\discretionary{.}{%
}{.}\hspace{.4pt}1111\discretionary{/}{%
}{/}jerd\hspace{.1pt}\discretionary{.}{%
}{.}\hspace{.4pt}13171}}}


\bibitem{martin2009head}
A.~Martin-Gonzalez, S.-M. Heining, and N.~Navab.
\newblock Head-mounted virtual loupe with sight-based activation for surgical applications.
\newblock In {\em 2009 8th IEEE international symposium on mixed and augmented reality}, pp. 207--208. IEEE, 2009. \href{https://doi.org/101109/ISMAR.2009.5336459}
{doi: {{%
101109\discretionary{/}{%
}{/}ISMAR\hspace{.1pt}\discretionary{.}{%
}{.}\hspace{.4pt}2009\hspace{.1pt}\discretionary{.}{%
}{.}\hspace{.4pt}5336459}}}


\bibitem{mendes2019survey}
D.~Mendes, F.~M. Caputo, A.~Giachetti, A.~Ferreira, and J.~Jorge.
\newblock A survey on 3d virtual object manipulation: From the desktop to immersive virtual environments.
\newblock In {\em Computer graphics forum}, vol.~38, pp. 21--45. Wiley Online Library, 2019. \href{https://doi.org/10.1111/cgf.13390}
{doi: {{%
10\hspace{.1pt}\discretionary{.}{%
}{.}\hspace{.4pt}1111\discretionary{/}{%
}{/}cgf\hspace{.1pt}\discretionary{.}{%
}{.}\hspace{.4pt}13390}}}


\bibitem{mendes2016benefits}
D.~Mendes, F.~Relvas, A.~Ferreira, and J.~Jorge.
\newblock The benefits of dof separation in mid-air 3d object manipulation.
\newblock In {\em Proceedings of the 22nd ACM conference on virtual reality software and technology}, pp. 261--268, 2016. \href{https://doi.org/10.1145/2993369.2993396}
{doi: {{%
10\hspace{.1pt}\discretionary{.}{%
}{.}\hspace{.4pt}1145\discretionary{/}{%
}{/}2993369\hspace{.1pt}\discretionary{.}{%
}{.}\hspace{.4pt}2993396}}}


\bibitem{mine1997moving}
M.~R. Mine, F.~P. Brooks~Jr, and C.~H. Sequin.
\newblock Moving objects in space: exploiting proprioception in virtual-environment interaction.
\newblock In {\em Proceedings of the 24th annual conference on Computer graphics and interactive techniques}, pp. 19--26, 1997. \href{https://doi.org/10.1145/258734.258747}
{doi: {{%
10\hspace{.1pt}\discretionary{.}{%
}{.}\hspace{.4pt}1145\discretionary{/}{%
}{/}258734\hspace{.1pt}\discretionary{.}{%
}{.}\hspace{.4pt}258747}}}


\bibitem{mistry20213d}
A.~Mistry, C.~Ucer, J.~D. Thompson, R.~S. Khan, E.~Karahmet, and F.~Sher.
\newblock 3d guided dental implant placement: impact on surgical accuracy and collateral damage to the inferior alveolar nerve.
\newblock {\em Dentistry Journal}, 9(9):99, 2021. \href{https://doi.org/10.3390/dj9090099}
{doi: {{%
10\hspace{.1pt}\discretionary{.}{%
}{.}\hspace{.4pt}3390\discretionary{/}{%
}{/}dj9090099}}}


\bibitem{muhler2009medical}
K.~Muhler, C.~Tietjen, F.~Ritter, and B.~Preim.
\newblock The medical exploration toolkit: An efficient support for visual computing in surgical planning and training.
\newblock {\em IEEE Transactions on Visualization and Computer Graphics}, 16(1):133--146, 2009. \href{https://doi.org/10.1109/TVCG.2009.58}
{doi: {{%
10\hspace{.1pt}\discretionary{.}{%
}{.}\hspace{.4pt}1109\discretionary{/}{%
}{/}TVCG\hspace{.1pt}\discretionary{.}{%
}{.}\hspace{.4pt}2009\hspace{.1pt}\discretionary{.}{%
}{.}\hspace{.4pt}58}}}


\bibitem{o2018use}
T.~O.~Nyumba, K.~Wilson, C.~J. Derrick, and N.~Mukherjee.
\newblock The use of focus group discussion methodology: Insights from two decades of application in conservation.
\newblock {\em Methods in Ecology and evolution}, 9(1):20--32, 2018. \href{https://doi.org/10.1111/2041-210X.12860}
{doi: {{%
10\hspace{.1pt}\discretionary{.}{%
}{.}\hspace{.4pt}1111\discretionary{/}{%
}{/}2041\discretionary{%
}{-}{-}210X\hspace{.1pt}\discretionary{.}{%
}{.}\hspace{.4pt}12860}}}


\bibitem{ohlendorf2017constrained}
D.~Ohlendorf, C.~Erbe, J.~Nowak, I.~Hauck, I.~Hermanns, D.~Ditchen, R.~Ellegast, and D.~A. Groneberg.
\newblock Constrained posture in dentistry--a kinematic analysis of dentists.
\newblock {\em BMC musculoskeletal disorders}, 18(1):1--15, 2017. \href{https://doi.org/10.1186/s12891-017-1650-x}
{doi: {{%
10\hspace{.1pt}\discretionary{.}{%
}{.}\hspace{.4pt}1186\discretionary{/}{%
}{/}s12891\discretionary{%
}{-}{-}017\discretionary{%
}{-}{-}1650\discretionary{%
}{-}{-}x}}}


\bibitem{park2015analysis}
H.-S. Park, J.~Kim, H.-L. Roh, and S.~Namkoong.
\newblock Analysis of the risk factors of musculoskeletal disease among dentists induced by work posture.
\newblock {\em Journal of physical therapy science}, 27(12):3651--3654, 2015. \href{https://doi.org/10.1589/jpts.27.3651}
{doi: {{%
10\hspace{.1pt}\discretionary{.}{%
}{.}\hspace{.4pt}1589\discretionary{/}{%
}{/}jpts\hspace{.1pt}\discretionary{.}{%
}{.}\hspace{.4pt}27\hspace{.1pt}\discretionary{.}{%
}{.}\hspace{.4pt}3651}}}


\bibitem{pellegrino2021dynamic}
G.~Pellegrino, A.~Ferri, M.~Del~Fabbro, C.~Prati, M.~Giovanna~Gandolfi, and C.~Marchetti.
\newblock Dynamic navigation in implant dentistry: A systematic review and meta-analysis.
\newblock {\em International Journal of Oral \& Maxillofacial Implants}, 36(5), 2021. \href{https://doi.org/10.11607/jomi.8770}
{doi: {{%
10\hspace{.1pt}\discretionary{.}{%
}{.}\hspace{.4pt}11607\discretionary{/}{%
}{/}jomi\hspace{.1pt}\discretionary{.}{%
}{.}\hspace{.4pt}8770}}}


\bibitem{pietruski2019supporting}
P.~Pietruski, M.~Majak, E.~{\'S}wiatek-Najwer, M.~{\.Z}uk, M.~Popek, M.~Mazurek, M.~{\'S}wiecka, and J.~Jaworowski.
\newblock Supporting mandibular resection with intraoperative navigation utilizing augmented reality technology--a proof of concept study.
\newblock {\em Journal of Cranio-Maxillofacial Surgery}, 47(6):854--859, 2019. \href{https://doi.org/10.1016/j.jcms.2019.03.004}
{doi: {{%
10\hspace{.1pt}\discretionary{.}{%
}{.}\hspace{.4pt}1016\discretionary{/}{%
}{/}j\hspace{.1pt}\discretionary{.}{%
}{.}\hspace{.4pt}jcms\hspace{.1pt}\discretionary{.}{%
}{.}\hspace{.4pt}2019\hspace{.1pt}\discretionary{.}{%
}{.}\hspace{.4pt}03\hspace{.1pt}\discretionary{.}{%
}{.}\hspace{.4pt}004}}}


\bibitem{preim2007visualization}
B.~Preim and D.~Bartz.
\newblock {\em Visualization in medicine: theory, algorithms, and applications}.
\newblock Elsevier, 2007.

\bibitem{qian2020ar}
L.~Qian, T.~Song, M.~Unberath, and P.~Kazanzides.
\newblock Ar-loupe: Magnified augmented reality by combining an optical see-through head-mounted display and a loupe.
\newblock {\em IEEE Transactions on Visualization and Computer Graphics}, 28(7):2550--2562, 2020. \href{https://doi.org/10.1109/TVCG.2020.3037284}
{doi: {{%
10\hspace{.1pt}\discretionary{.}{%
}{.}\hspace{.4pt}1109\discretionary{/}{%
}{/}TVCG\hspace{.1pt}\discretionary{.}{%
}{.}\hspace{.4pt}2020\hspace{.1pt}\discretionary{.}{%
}{.}\hspace{.4pt}3037284}}}


\bibitem{raikar2017factors}
S.~Raikar, P.~Talukdar, S.~Kumari, S.~K. Panda, V.~M. Oommen, and A.~Prasad.
\newblock Factors affecting the survival rate of dental implants: A retrospective study.
\newblock {\em Journal of International Society of Preventive \& Community Dentistry}, 7(6):351, 2017. \href{https://doi.org/10.4103/jispcd.JISPCD_380_17}
{doi: {{%
10\hspace{.1pt}\discretionary{.}{%
}{.}\hspace{.4pt}4103\discretionary{/}{%
}{/}jispcd\hspace{.1pt}\discretionary{.}{%
}{.}\hspace{.4pt}JISPCD\_380\_17}}}


\bibitem{schneider2009systematic}
D.~Schneider, P.~Marquardt, M.~Zwahlen, and R.~E. Jung.
\newblock A systematic review on the accuracy and the clinical outcome of computer-guided template-based implant dentistry.
\newblock {\em Clinical oral implants research}, 20:73--86, 2009. \href{https://doi.org/10.1111/j.1600-0501.2009.01788.x}
{doi: {{%
10\hspace{.1pt}\discretionary{.}{%
}{.}\hspace{.4pt}1111\discretionary{/}{%
}{/}j\hspace{.1pt}\discretionary{.}{%
}{.}\hspace{.4pt}1600\discretionary{%
}{-}{-}0501\hspace{.1pt}\discretionary{.}{%
}{.}\hspace{.4pt}2009\hspace{.1pt}\discretionary{.}{%
}{.}\hspace{.4pt}01788\hspace{.1pt}\discretionary{.}{%
}{.}\hspace{.4pt}x}}}


\bibitem{schubert2019digital}
O.~Schubert, J.~Schweiger, M.~Stimmelmayr, E.~Nold, and J.-F. G{\"u}th.
\newblock Digital implant planning and guided implant surgery--workflow and reliability.
\newblock {\em British dental journal}, 226(2):101--108, 2019. \href{https://doi.org/10.1038/sj.bdj.2019.44}
{doi: {{%
10\hspace{.1pt}\discretionary{.}{%
}{.}\hspace{.4pt}1038\discretionary{/}{%
}{/}sj\hspace{.1pt}\discretionary{.}{%
}{.}\hspace{.4pt}bdj\hspace{.1pt}\discretionary{.}{%
}{.}\hspace{.4pt}2019\hspace{.1pt}\discretionary{.}{%
}{.}\hspace{.4pt}44}}}


\bibitem{shaalan2020visualization}
D.~Shaalan and S.~Jusoh.
\newblock Visualization in medical system interfaces: Ux guidelines.
\newblock In {\em 2020 12th International Conference on Electronics, Computers and Artificial Intelligence (ECAI)}, pp. 1--8. IEEE, 2020. \href{https://doi.org/10.1109/ECAI50035.2020.9223236}
{doi: {{%
10\hspace{.1pt}\discretionary{.}{%
}{.}\hspace{.4pt}1109\discretionary{/}{%
}{/}ECAI50035\hspace{.1pt}\discretionary{.}{%
}{.}\hspace{.4pt}2020\hspace{.1pt}\discretionary{.}{%
}{.}\hspace{.4pt}9223236}}}


\bibitem{slattery2020research}
P.~Slattery, A.~K. Saeri, and P.~Bragge.
\newblock Research co-design in health: a rapid overview of reviews.
\newblock {\em Health research policy and systems}, 18(1):1--13, 2020. \href{https://doi.org/10.1186/s12961-020-0528-9}
{doi: {{%
10\hspace{.1pt}\discretionary{.}{%
}{.}\hspace{.4pt}1186\discretionary{/}{%
}{/}s12961\discretionary{%
}{-}{-}020\discretionary{%
}{-}{-}0528\discretionary{%
}{-}{-}9}}}


\bibitem{song2018endodontic}
T.~Song, C.~Yang, O.~Dianat, and E.~Azimi.
\newblock Endodontic guided treatment using augmented reality on a head-mounted display system.
\newblock {\em Healthcare Technology Letters}, 5(5):201--207, 2018. \href{https://doi.org/10.1049/htl.2018.5062}
{doi: {{%
10\hspace{.1pt}\discretionary{.}{%
}{.}\hspace{.4pt}1049\discretionary{/}{%
}{/}htl\hspace{.1pt}\discretionary{.}{%
}{.}\hspace{.4pt}2018\hspace{.1pt}\discretionary{.}{%
}{.}\hspace{.4pt}5062}}}


\bibitem{spille2022learning}
J.~Spille, E.~Helmstetter, P.~K{\"u}bel, J.-T. Weitkamp, J.~Wagner, H.~Wieker, H.~Naujokat, C.~Fl{\"o}rke, J.~Wiltfang, and A.~G{\"u}lses.
\newblock Learning curve and comparison of dynamic implant placement accuracy using a navigation system in young professionals.
\newblock {\em Dentistry Journal}, 10(10):187, 2022. \href{https://doi.org/10.3390/dj10100187}
{doi: {{%
10\hspace{.1pt}\discretionary{.}{%
}{.}\hspace{.4pt}3390\discretionary{/}{%
}{/}dj10100187}}}


\bibitem{takacs2023advancing}
A.~Takacs, E.~Hardi, B.~G.~N. Cavalcante, B.~Szabo, B.~Kispelyi, A.~Joob-Fancsaly, K.~Mikulas, G.~Varga, P.~Hegyi, and M.~Kivovics.
\newblock Advancing accuracy in guided implant placement: A comprehensive meta-analysis: Meta-analysis evaluation of the accuracy of available implant placement methods.
\newblock {\em Journal of Dentistry}, 139:104748, 2023. \href{https://doi.org/10.1016/j.jdent.2023.104748}
{doi: {{%
10\hspace{.1pt}\discretionary{.}{%
}{.}\hspace{.4pt}1016\discretionary{/}{%
}{/}j\hspace{.1pt}\discretionary{.}{%
}{.}\hspace{.4pt}jdent\hspace{.1pt}\discretionary{.}{%
}{.}\hspace{.4pt}2023\hspace{.1pt}\discretionary{.}{%
}{.}\hspace{.4pt}104748}}}


\bibitem{tao2023comparison}
B.~Tao, X.~Fan, F.~Wang, X.~Chen, Y.~Shen, and Y.~Wu.
\newblock Comparison of the accuracy of dental implant placement using dynamic and augmented reality-based dynamic navigation: An in vitro study.
\newblock {\em Journal of Dental Sciences}, 2023. \href{https://doi.org/10.1016/j.jds.2023.05.006}
{doi: {{%
10\hspace{.1pt}\discretionary{.}{%
}{.}\hspace{.4pt}1016\discretionary{/}{%
}{/}j\hspace{.1pt}\discretionary{.}{%
}{.}\hspace{.4pt}jds\hspace{.1pt}\discretionary{.}{%
}{.}\hspace{.4pt}2023\hspace{.1pt}\discretionary{.}{%
}{.}\hspace{.4pt}05\hspace{.1pt}\discretionary{.}{%
}{.}\hspace{.4pt}006}}}


\bibitem{tao2024comparison}
B.~Tao, X.~Fan, F.~Wang, X.~Chen, Y.~Shen, and Y.~Wu.
\newblock Comparison of the accuracy of dental implant placement using dynamic and augmented reality-based dynamic navigation: an in vitro study.
\newblock {\em Journal of Dental Sciences}, 19(1):196--202, 2024. \href{https://doi.org/10.1016/j.jds.2023.05.006}
{doi: {{%
10\hspace{.1pt}\discretionary{.}{%
}{.}\hspace{.4pt}1016\discretionary{/}{%
}{/}j\hspace{.1pt}\discretionary{.}{%
}{.}\hspace{.4pt}jds\hspace{.1pt}\discretionary{.}{%
}{.}\hspace{.4pt}2023\hspace{.1pt}\discretionary{.}{%
}{.}\hspace{.4pt}05\hspace{.1pt}\discretionary{.}{%
}{.}\hspace{.4pt}006}}}


\bibitem{Unity}
Unity.
\newblock {\em Unity}.
\newblock last accessed on 10/01/2024.

\bibitem{volmer2023multi}
B.~Volmer, J.-S. Liu, B.~Matthews, I.~Bornkessel-Schlesewsky, S.~Feiner, and B.~H. Thomas.
\newblock Multi-level precues for guiding tasks within and between workspaces in spatial augmented reality.
\newblock {\em IEEE Transactions on Visualization and Computer Graphics}, 29(11):4449--4459, 2023. \href{https://doi.org/10.1109/TVCG.2018.2868587}
{doi: {{%
10\hspace{.1pt}\discretionary{.}{%
}{.}\hspace{.4pt}1109\discretionary{/}{%
}{/}TVCG\hspace{.1pt}\discretionary{.}{%
}{.}\hspace{.4pt}2018\hspace{.1pt}\discretionary{.}{%
}{.}\hspace{.4pt}2868587}}}


\bibitem{waelkens2016surgical}
P.~Waelkens, M.~N. van Oosterom, N.~S. van~den Berg, N.~Navab, and F.~W. van Leeuwen.
\newblock Surgical navigation: an overview of the state-of-the-art clinical applications.
\newblock {\em Radioguided Surgery: Current Applications and Innovative Directions in Clinical Practice}, pp. 57--73, 2016. \href{https://doi.org/10.1007/978-3-319-26051-8_4}
{doi: {{%
10\hspace{.1pt}\discretionary{.}{%
}{.}\hspace{.4pt}1007\discretionary{/}{%
}{/}978\discretionary{%
}{-}{-}3\discretionary{%
}{-}{-}319\discretionary{%
}{-}{-}26051\discretionary{%
}{-}{-}8\_4}}}


\bibitem{wajngarten2021magnification}
D.~Wajngarten, J.~M. Pazos, V.~P. Menegazzo, J.~P.~D. Novo, and P.~P. N.~S. Garcia.
\newblock Magnification effect on fine motor skills of dental students.
\newblock {\em Plos one}, 16(11):e0259768, 2021. \href{https://doi.org/10.1371/journal.pone.0259768}
{doi: {{%
10\hspace{.1pt}\discretionary{.}{%
}{.}\hspace{.4pt}1371\discretionary{/}{%
}{/}journal\hspace{.1pt}\discretionary{.}{%
}{.}\hspace{.4pt}pone\hspace{.1pt}\discretionary{.}{%
}{.}\hspace{.4pt}0259768}}}


\bibitem{wang2014augmented}
J.~Wang, H.~Suenaga, K.~Hoshi, L.~Yang, E.~Kobayashi, I.~Sakuma, and H.~Liao.
\newblock Augmented reality navigation with automatic marker-free image registration using 3-d image overlay for dental surgery.
\newblock {\em IEEE transactions on biomedical engineering}, 61(4):1295--1304, 2014. \href{https://doi.org/10.1109/TBME.2014.2301191}
{doi: {{%
10\hspace{.1pt}\discretionary{.}{%
}{.}\hspace{.4pt}1109\discretionary{/}{%
}{/}TBME\hspace{.1pt}\discretionary{.}{%
}{.}\hspace{.4pt}2014\hspace{.1pt}\discretionary{.}{%
}{.}\hspace{.4pt}2301191}}}


\bibitem{wang2023exploring}
W.~Wang, M.~Zhuang, S.~Li, Y.~Shen, R.~Lan, Y.~Wu, and F.~Wang.
\newblock Exploring training dental implant placement using static or dynamic devices among dental students.
\newblock {\em European Journal of Dental Education}, 27(3):438--448, 2023. \href{https://doi.org/10.1111/eje.12825}
{doi: {{%
10\hspace{.1pt}\discretionary{.}{%
}{.}\hspace{.4pt}1111\discretionary{/}{%
}{/}eje\hspace{.1pt}\discretionary{.}{%
}{.}\hspace{.4pt}12825}}}


\bibitem{wu2024impacts}
B.-Z. Wu and F.~Sun.
\newblock The impacts of registration-and-fixation device positioning on the performance of implant placement assisted by dynamic computer-aided surgery: A randomized controlled trial.
\newblock {\em Clinical Oral Implants Research}, 2024. \href{https://doi.org/10.1111/clr.14237}
{doi: {{%
10\hspace{.1pt}\discretionary{.}{%
}{.}\hspace{.4pt}1111\discretionary{/}{%
}{/}clr\hspace{.1pt}\discretionary{.}{%
}{.}\hspace{.4pt}14237}}}


\bibitem{yari2024risk}
A.~Yari, P.~Fasih, S.~Alborzi, H.~Nikzad, and E.~Romoozi.
\newblock Risk factors associated with early implant failure: A retrospective review.
\newblock {\em Journal of Stomatology, Oral and Maxillofacial Surgery}, 125(4):101749, 2024. \href{https://doi.org/10.1016/j.jormas.2023.101749}
{doi: {{%
10\hspace{.1pt}\discretionary{.}{%
}{.}\hspace{.4pt}1016\discretionary{/}{%
}{/}j\hspace{.1pt}\discretionary{.}{%
}{.}\hspace{.4pt}jormas\hspace{.1pt}\discretionary{.}{%
}{.}\hspace{.4pt}2023\hspace{.1pt}\discretionary{.}{%
}{.}\hspace{.4pt}101749}}}


\bibitem{yeo2024entering}
A.~Yeo, B.~W. Kwok, A.~Joshna, K.~Chen, and J.~S. Lee.
\newblock Entering the next dimension: A review of 3d user interfaces for virtual reality.
\newblock {\em Electronics}, 13(3):600, 2024. \href{https://doi.org/10.3390/electronics13030600}
{doi: {{%
10\hspace{.1pt}\discretionary{.}{%
}{.}\hspace{.4pt}3390\discretionary{/}{%
}{/}electronics13030600}}}


\bibitem{zhai1993human}
S.~Zhai and P.~Milgram.
\newblock Human performance evaluation of manipulation schemes in virtual environments.
\newblock In {\em Proceedings of IEEE Virtual Reality Annual International Symposium}, pp. 155--161. IEEE, 1993. \href{https://doi.org/10.1109/VRAIS.1993.380784}
{doi: {{%
10\hspace{.1pt}\discretionary{.}{%
}{.}\hspace{.4pt}1109\discretionary{/}{%
}{/}VRAIS\hspace{.1pt}\discretionary{.}{%
}{.}\hspace{.4pt}1993\hspace{.1pt}\discretionary{.}{%
}{.}\hspace{.4pt}380784}}}


\bibitem{zhu2017novel}
M.~Zhu, F.~Liu, G.~Chai, J.~J. Pan, T.~Jiang, L.~Lin, Y.~Xin, Y.~Zhang, and Q.~Li.
\newblock A novel augmented reality system for displaying inferior alveolar nerve bundles in maxillofacial surgery.
\newblock {\em Scientific Reports}, 7(1):42365, 2017. \href{https://doi.org/10.1038/srep42365}
{doi: {{%
10\hspace{.1pt}\discretionary{.}{%
}{.}\hspace{.4pt}1038\discretionary{/}{%
}{/}srep42365}}}


\end{thebibliography}

\end{document}